\documentclass[prd,twocolumn,showpacs,floatfix,amsmath,nofootinbib,amssymb,floatfix]{revtex4}
\usepackage{graphicx,color,dcolumn,booktabs,bm}
\usepackage{longtable,lscape}
\usepackage{txfonts}
\usepackage{overpic}
\usepackage{amssymb}
\usepackage{indentfirst}
\usepackage{feynmf}   
\usepackage{slashed}  
\usepackage{cases}
\usepackage{color}
\usepackage{multirow}
\usepackage{epstopdf}
\usepackage{graphicx,color,dcolumn,booktabs,bm}
\usepackage[colorlinks,
            citecolor=blue,
            anchorcolor=red,
            menucolor=red,
            linkcolor=red,
            filecolor=red,
            runcolor=red,
            urlcolor=blue,
            frenchlinks=red]{hyperref}

\graphicspath{{Figures/}} %

\allowdisplaybreaks

\begin{document}
\title{Triply heavy tetraquark states with the $QQ\bar{Q}\bar{q}$ configuration}
\author{Kan Chen$^{1,2}$}
\author{Xiang Liu$^{1,2}$}\email{xiangliu@lzu.edu.cn}
\author{Jing Wu$^{3}$}
\author{Yan-Rui Liu$^{3}$}\email{yrliu@sdu.edu.cn}
\author{Shi-Lin Zhu$^{4,5,6}$}\email{zhusl@pku.edu.cn}
%
%
\address{$^1$School of Physical Science and Technology, Lanzhou
University, Lanzhou 730000, China\\
$^2$Research Center for Hadron and CSR Physics, Lanzhou University and Institute of Modern Physics of CAS, Lanzhou 730000, China\\
$^3$School of Physics and Key Laboratory of Particle Physics and Particle Irradiation (MOE), Shandong University, Jinan 250100, China \\
$^4$School of Physics and State Key Laboratory of Nuclear Physics and Technology, Peking University, Beijing 100871, China \\
$^5$Collaborative Innovation Center of Quantum Matter, Beijing 100871, China\\
$^6$Center of High Energy Physics, Peking University, Beijing 100871, China
}
%
%
\begin{abstract}{ In the framework of the color-magnetic interaction, we
systematically investigate the mass splittings of the
$QQ\bar{Q}\bar{q}$ tetraquark states and estimated their rough
masses in this work. These systems include the explicitly exotic
states $cc\bar{b}\bar{q}$ and $bb\bar{c}\bar{q}$ and the hidden
exotic states $cc\bar{c}\bar{q}$, $cb\bar{b}\bar{q}$,
$bc\bar{c}\bar{q}$, and $bb\bar{b}\bar{q}$. If a state around the
estimated mass region could be observed, its nature as a genuine
tetraquark is favored. The strong decay patterns shown here will be
helpful to the experimental search for these exotic states.}
\end{abstract}

\maketitle
%
\section{Introduction}\label{sec1}

In the past decade, many exotic $XYZ$ states were observed in
experiments
\cite{Swanson:2006st,Zhu:2007wz,Voloshin:2007dx,Drenska:2010kg,Esposito:2014rxa,Chen:2016qju,Chen:2016heh,Hosaka:2016pey,Richard:2016eis,Lebed:2016hpi,Esposito:2016noz}.
Among them, the charged charmonium-like or bottomonium-like states,
the $Z(4430)$
\cite{Choi:2007wga,Mizuk:2009da,Chilikin:2014bkk,Aaij:2014jqa}, the
$Z_1(4050)$ \cite{Mizuk:2008me}, the $Z_2(4250)$
\cite{Mizuk:2008me}, the $Z_c(3900)$
\cite{Ablikim:2013mio,Liu:2013dau,Xiao:2013iha,Ablikim:2015tbp}, the
$Z_c(3885)$ \cite{Ablikim:2013xfr,Ablikim:2015swa,Ablikim:2015gda},
the $Z_c(4020)$\cite{Ablikim:2013wzq,Ablikim:2014dxl}, the
$Z_c(4025) $\cite{Ablikim:2013emm,Ablikim:2015vvn}, the $Z_c(4200)
$\cite{Chilikin:2014bkk}, the $Z_b(10610)$ \cite{Belle:2011aa} and
the $Z_{b} $(10650) \cite{Belle:2011aa}, are considered as possible
tetraquatk candidates with two heavy quarks. They are also good
meson-antimeson molecule candidates.

Besides the hidden flavor case, the exotic charmed mesons also
stimulated heated discussions on tetraquark candidates. The possible
candidates of the exotic charmed mesons $D_s(2317)$
\cite{Aubert:2003fg,Besson:2003cp,Krokovny:2003zq}, $D_s(2460)$
\cite{Besson:2003cp,Krokovny:2003zq} and $D_s(2632)$
\cite{Evdokimov:2004iy} have attracted much attention due to the
deviation of their masses from the quark model expectations
\cite{Godfrey:1985xj} and their unexpected decay properties. Cheng
and Hou \cite{Cheng:2003kg} interpreted the $D_{s}(2317)$ as a
$cq\bar{q}\bar{s}$ state to explain its mass and decay behaviors. In
Ref. \cite{Chen:2004dy}, Chen and Li proposed that the $D_s(2317)$
($D_s(2460)$) is a $DK$ ($D^*K$) molecule while the $D_s(2632)$ is a
$c\bar{s}s\bar{s}$ state. The decay processes $D_s(2317)\to
D_s^+\pi^0, D_s^{*+}\gamma$ are studied in a four-quark meson
assumption in Ref. \cite{Hayashigaki:2004st}, which favors the
assignment as an iso-triplet. Maiani et al. suggested the
$D_s(2632)$ as a $[cd\bar{d}\bar{s}]$ state \cite{Maiani:2004xg} to
explain that the $D^0K^+$ mode is suppressed with respect to the
$D_{s}^+\eta$ channel. On the other hand, Liu et al.
\cite{Liu:2004kd} proposed that the anomalous decay ratio of the
$D_s(2632)$ can be understood by assuming a four quark state wave
function
$\frac{1}{2\sqrt{2}}(ds\bar{d}+sd\bar{d}+su\bar{u}+us\bar{u}-2ss\bar{s})\bar{c}$.
The $D_s(2632)$ was not confirmed later. The $D_s(2317)$ and the
$D_s(2460)$ mesons are probably conventional charm-strange mesons
which are affected largely by the coupled channel effects
\cite{Lang:2014yfa}. The existence of the open flavor tetraquarks
remains elusive.

Recently, the D\O\, Collaboration \cite{D0:2016mwd} reported a
structure $X(5568)$ in the $B_s^0\pi^{\pm}$ invariant mass
distribution. This narrow state is about 200 MeV below the
$B\bar{K}$ threshold. Thus, the $X(5568)$ could be a tetraquark
state with four different flavors
\cite{Agaev:2016mjb,Wang:2016mee,Wang:2016tsi,Chen:2016mqt,Zanetti:2016wjn,Agaev:2016ijz,Liu:2016ogz,Agaev:2016lkl,Dias:2016dme,Wang:2016wkj,He:2016yhd,Stancu:2016sfd,Tang:2016pcf,Goerke:2016hxf}
rather than a molecule state because the binding is too deep for a
weak $B\bar{K}$ interaction in the isovector case
\cite{Liu:2009uz,Agaev:2016urs,Chen:2016ypj,Lu:2016kxm}. Later, the
LHCb Collaboration also investigated the $X(5568)$ state
\cite{Aaij:2016iev}, but found no significant signals, which led
some theorists to doubt whether the $X(5568)$ is a genuine resonance
\cite{Liu:2016xly,Lu:2016zhe,Esposito:2016itg,Albaladejo:2016eps,Ali:2016gdg,Albuquerque:2016nlw,Chen:2016npt,Jin:2016cpv,Burns:2016gvy,Guo:2016nhb,Kang:2016zmv}.
The CMS Collaboration did not confirm the state, either
\cite{CMS5568}. In this work, we move on to the four quark systems
with more heavy quarks to search for more compact tetraquark states,
such as $QQ\bar{Q}\bar{Q}$.

In Ref. \cite{Iwasaki:1975pv}, a structure composed of a
charm-anticharm quark pair as well as a light quark-antiquark pair,
$c\bar{c}q\bar{q}$, was proposed. Other exotic mesons with hidden
charm including $c\bar{c}c\bar{c}$ were also discussed. After that,
more groups discussed the existence of such exotic states in various
methods
\cite{Chao:1980dv,Ader:1981db,Heller:1985cb,SilvestreBrac:1992mv,SilvestreBrac:1993ss,Lloyd:2003yc,Barnea:2006sd,Wu:2016vtq}.
There are different opinions on the stability of the
$c\bar{c}c\bar{c}$ system. Whether such superheavy tetraquark states
exist or not awaits experimental judgement in the future.

Usually, it is difficult for one to distinguish a meson-antimeson
molecule from a compact tetraquark. However, for the
$QQ\bar{Q}\bar{Q}$ system, the binding force comes from the
short-range gluon exchange and a molecule configuration is not
favored. If such states do exist, it is also possible that the
compact tetraquarks with three heavy quarks and one light quark
$QQ\bar{Q}\bar{q}$ may exist. The binding force is also provided by
the short-range gluon exchange. These states look like the excited
$D$ or $B$ mesons.

Recently, two excited nucleon states $P_c(4380)$ and $P_c$(4450)
were observed by the LHCb Collaboration
\cite{Aaij:2015tga,Aaij:2016phn,Aaij:2016ymb}, where a
charm-anticharm pair is excited. Such pentaquarks were predicted in
the baryon-meson picture in Refs.
\cite{Wu:2010jy,Wu:2010vk,Yang:2011wz,Wang:2011rga}. It is also
possible that a $c\bar{c}$ or $b\bar{b}$ pair may be excited in the
charmed or bottomed mesons. Once such states are really observed,
their molecule assignment is not supported while the tetraquark
nature is favorable. The experimental and theoretical search for
this kind of tetraquark states is missing, as far as we know. In
this work, we perform a systematic analysis of the mass spectrum of
the $QQ\bar{Q}\bar{q}$ system, which may provide important
information for future experimental research.

Because the interaction strengths between (anti)quarks may be
different, it is usually assumed that the diquark substructure
exists in multiquark states, which is reflected in the mass spectra.
In studying pentaquark states \cite{Karliner:2003dt}, it was argued
that the triquark ($qq\bar{q}$) substructure results in a lower
hadron mass. If one specifies the substructure with a given
color-spin state, the two configurations result in different
spectra. Here we consider all possible color-spin states and finally
diagonalize the Hamiltonian. To understand whether they are
equivalent for the compact $QQ\bar{Q}\bar{q}$ system, we estimate
the mass in the diquark-antidiquark $(QQ)(\bar{Q}\bar{q})$ picture
and the triquark-antiquark $(QQ\bar{Q})\bar{q}$ picture. One will
see that, for the present systems, the two configurations give the
same results once the diagonalization is performed.

This paper is organized as follows. In Sec. \ref{sec2}, we present
the formalism of our calculation. In Sec. \ref{sec3}, we show the
numerical results for the various $QQ\bar{Q}\bar{q}$ systems.
Finally, we give some discussions and a short summary in the last
section.

\section{Formalism}\label{sec2}

In Ref. \cite{DeRujula:1975qlm}, de Rujula, Georgi, and Glashow
proposed a nonrelativistic Hamiltonian which consists of the
one-gluon exchange potential and a non-perturbative scalar confining
potential. For the $S$-wave ground states, the spin-orbit and the
tensor interactions may be ignored. Then, the Hamiltonian has the
form
\begin{eqnarray}\label{FB}
H&=&\sum_i(m_i+\frac{p_i^2}{2m_i})+\sum_{i<j}V_1(r_{ij})\lambda_i\cdot\lambda_j\nonumber\\
&&+\sum_{i<j}V_2(r_{ij})\lambda_i\cdot\lambda_j\sigma_i\cdot\sigma_j,
\end{eqnarray}
where $m_i$ is the constituent quark mass of the $i$-th quark, $p_i$
is the three-momentum of the $i$-th quark, $r_{ij}$ is the distance
between the quarks labeled with $i$ and $j$, and
$\sigma_{i}^{x,y,z}$ ($\lambda_{i}^{1,2,\cdots,8}$) are the Pauli
(Gell-Mann) matrices corresponding to the $i$-th quark. For
antiquarks, $\lambda_i$ should be replaced by $-\lambda^*_i$. The
potential $V_1(r)$ includes the color-coulomb, color-electric, and
confinement terms while the potential $V_2(r)$ is a short-range
contact term. The resulting eigenvalues of this Hamiltonian are the
masses of the two-quark mesons or three-quark baryons.

According to this model, the mass splittings between the ground
state hadrons with the same quark content are determined mainly by
the color-spin (chromomagnetic or color-magnetic) interaction. For
example, this interaction term accounts for the mass difference
between the nucleon and the $\Delta$ resonance. When we mainly focus
on the mass splittings, one may reduce the above formula to the
following simple form after the average in the coordinate space is
taken,
\begin{eqnarray}\label{Hcmi}
H&=&H_{0}+H_{CM}\nonumber\\
&=&\sum_im_i^{eff}-\sum_{i<j}C_{ij}\lambda_i\cdot\lambda_j\sigma_i\cdot\sigma_j,
\end{eqnarray}
where $m_i^{eff}$ is the effective mass of the $i$-th quark which
includes the constituent quark mass and the binding effects. The
coupling constant $C_{ij}=c_0\langle\delta(r_{ij})\rangle/(m_im_j)$,
which is a real number, is determined by the wave function and
the constituent quark mass, where $c_0$ is related to the
interaction constant. The reduction may be performed because
$\langle \lambda_i\cdot\lambda_j\rangle=-\frac{16}{3}$ for all the
two-quark mesons and $-\frac83$ for all the three-quark baryons. In
principle, the value of an effective quark mass in one system may be
different from that in another system.

In the study of $S$-wave compact multiquark states, the average of
$\langle\lambda_i\cdot\lambda_j\rangle$ depending on the color
structure of the two quarks or the quark-antiquark pair may be
$-\frac83$ for $\bar{3}_c$, $\frac43$ for $6_c$, $-\frac{16}{3}$ for
$1_c$, or $\frac23$ for $8_c$. Probably all these numbers contribute
to a multiquark state with a given quark content and the above
reduction might seem problematic at first sight. However, the
multiquark hadron must be a color singlet state and its color wave
function is a linear superposition of different color
configurations. The distribution of the color configurations is
roughly determined by the kinetic term and the $V_1$ term in Eq.
(\ref{FB}), which defines the multiquark multiplet. Once the color
distribution or the full wave function of the multiquark state is
known, one may further get the mass splittings with the $V_2$ term.
In other words, the above reduced Hamiltonian is also applicable to
the multiquark systems. Various investigations with this simple
model in studying the mass spectra of multiquark states may be found
in the literature, e.g.
\cite{Wu:2016vtq,Maiani:2004vq,Hogaasen:2005jv,Stancu:2009ka,Kim:2016tys,Wu:2016gas}. For
the present $Q_1Q_2\bar{Q}_3\bar{q}_4$ system, the explicit CMI
(color-magnetic interaction) term is
\begin{eqnarray}
H_{CM}&=&-C_{12}\lambda_1\cdot\lambda_2\,\sigma_1\cdot\sigma_2-C_{34}\lambda_3^*\cdot\lambda_4^*\,\sigma_3\cdot\sigma_4\nonumber\\
&&+\sum_{i=1,2;j=3,4}C_{ij}\lambda_i\cdot\lambda_j^*\sigma_i\cdot\sigma_j.
\end{eqnarray}

In order to calculate the color-spin matrix elements, we adopt the
formalism in Ref. \cite{Hogaasen:2004pm,Buccella:2006fn}. First, we
construct the color and spin wave functions by using both the
diquark-antidiquark and the triquark-antiquark configurations, and
then calculate the color and spin matrix elements with the
Hamiltonians $H_C=-\sum_{i<j}C_{ij}\lambda_i\cdot\lambda_j$ and
$H_S=-\sum_{i<j}C_{ij}\sigma_i\cdot\sigma_j$, respectively. Finally,
we get the $\langle H_{CM}\rangle$ matrices after performing a type
of ``tensor product'' of $\langle H_C\rangle$ and $\langle
H_S\rangle$.

We take the calculation of the matrix element $\langle
\phi_\alpha\chi_x|H_{CM}|\phi_\beta\chi_y\rangle$ as an example to
illustrate the meaning of the ``tensor product''. Here,
$\phi_{\alpha}$ or $\phi_{\beta}$ ($\chi_{x}$ or $\chi_{y}$)
indicates the color (spin) wave function of a tetraquark system. If
one has obtained $\langle
\chi_x|H_S|\chi_y\rangle=a_SC_{12}+b_SC_{13}+\cdots$ and
$\langle\phi_\alpha|H_C|\phi_\beta\rangle$
$=a_CC_{12}+b_CC_{13}+\cdots$, one gets $\langle
\phi_\alpha\chi_x|H_{CM}|\phi_\beta\chi_y\rangle$
$=-[(a_S*a_C)C_{12}+(b_S*b_C)C_{13}+\cdots]$. 

In the diquark-antidiquark configuration, the allowed base vectors
in spin space read
\begin{eqnarray}
&\chi_1=|(Q_1Q_2)_1(\bar{Q}_3\bar{q}_4)_1\rangle_2,\quad \chi_2=|(Q_1Q_2)_1(\bar{Q}_3\bar{q}_4)_1\rangle_1,\nonumber\\
&\chi_3=|(Q_1Q_2)_1(\bar{Q}_3\bar{q}_4)_1\rangle_0,\quad \chi_4=|(Q_1Q_2)_1(\bar{Q}_3\bar{q}_4)_0\rangle_1,\nonumber\\
&\chi_5=|(Q_1Q_2)_0(\bar{Q}_3\bar{q}_4)_1\rangle_1,\quad
\chi_6=|(Q_1Q_2)_0(\bar{Q}_3\bar{q}_4)_0\rangle_0,
\end{eqnarray}
where the notation on the right hand side is $|(Q_1Q_2)_{spin}$
$(\bar{Q}_3\bar{q}_4)_{spin}\rangle_{spin}$. Similarly, the base
vectors in spin space in the triquark-antiquark configuration are
\begin{eqnarray}
&\zeta_1=|[(Q_1Q_2)_1\bar{Q}_3]_\frac{3}{2}\bar{q}_4\rangle_2,\quad \zeta_2=|[(Q_1Q_2)_1\bar{Q}_3]_\frac{3}{2}\bar{q}_4\rangle_1,\nonumber\\
&\zeta_3=|[(Q_1Q_2)_1\bar{Q}_3]_\frac{1}{2}\bar{q}_4\rangle_1,\quad \zeta_4=|[(Q_1Q_2)_1\bar{Q}_3]_\frac{1}{2}\bar{q}_4\rangle_0,\nonumber\\
&\zeta_5=|[(Q_1Q_2)_0\bar{Q}_3]_\frac{1}{2}\bar{q}_4\rangle_1,\quad
\zeta_6=|[(Q_1Q_2)_0\bar{Q}_3]_\frac{1}{2}\bar{q}_4\rangle_0,
\end{eqnarray}
where the notation
$|[(Q_1Q_2)_{spin}\bar{Q}_3]_{spin}\bar{q}_4)\rangle_{spin}$ is
used. With the explicit spin wave functions, one finds that
$\chi_1=\zeta_1$, $\chi_3=\zeta_4$, $\chi_5=\zeta_5$, and
$\chi_6=\zeta_6$.

The color wave functions for the diquark belong to the $6_c$ or
$\bar{3}_c$ representation, while those of the anti-diquark belong
to the $\bar{6}_c$ or $3_c$ representation. In the
diquark-antidiquark configuration, the base vectors in color space
are
\begin{equation}
\phi_1=|(Q_1Q_2)^6(\bar{Q}_3\bar{q}_4)^{\bar{6}}\rangle,\quad
\phi_2=|(Q_1Q_2)^{\bar{3}}(\bar{Q}_3\bar{q}_4)^{3}\rangle,
\end{equation}
where the superscripts are color representations. In the
triquark-antiquark configuration, there are two triplet
representations $3_{MS}$ and $3_{MA}$ for the triquark where $MS$
($MA$) means that the first two quarks are symmetric
(antisymmetric). The base vectors
$|(Q_1Q_2\bar{Q}_3)^{3_{MS}}\bar{q}_4\rangle$ and $|(Q_1Q_2$
$\bar{Q}_3)^{3_{MA}}\bar{q}_4\rangle$ seem different from those in
the diquark-anti-diquark configuration. However, we find that the
two configurations give the same results by constructing the
explicit color wave functions. That is,
\begin{eqnarray}
\phi_1&=&|(Q_1Q_2)^6(\bar{Q}_3\bar{q}_4)^{\bar{6}}\rangle=|(Q_1Q_2\bar{Q}_3)^{3_{MS}}\bar{q}_4\rangle\nonumber\\
&=&\frac{1}{2\sqrt{6}}[2(bb\bar{b}\bar{b}+rr\bar{r}\bar{r}+gg\bar{g}\bar{g})+rb\bar{b}\bar{r}
+rb\bar{r}\bar{b}+br\bar{b}\bar{r}\nonumber\\
&&+br\bar{r}\bar{b}+gb\bar{b}\bar{g}+
gb\bar{g}\bar{b}+bg\bar{b}\bar{g}+bg\bar{g}\bar{b}+gr\bar{r}\bar{g}+
gr\bar{g}\bar{r}\nonumber\\
&&+rg\bar{g}\bar{r}+rg\bar{r}\bar{g}],\nonumber\\
\phi_2&=&|(Q_1Q_2)^{\bar{3}}(\bar{Q}_3\bar{q}_4)^{3}\rangle=|(Q_1Q_2\bar{Q}_3)^{3_{MA}}\bar{q}_4\rangle\rangle\nonumber\\
&=&\frac{1}{2\sqrt{3}}[gr\bar{r}\bar{g}-gr\bar{g}\bar{r}-rg\bar{r}\bar{g}+rg\bar{g}\bar{r}+
rb\bar{b}\bar{r}-rb\bar{r}\bar{b}\nonumber\\
&&-br\bar{b}\bar{r}+br\bar{r}\bar{b}-
gb\bar{g}\bar{b}+gb\bar{b}\bar{g}+bg\bar{g}\bar{b}-bg\bar{b}\bar{g}].
\end{eqnarray}

For the $color\otimes spin$ wave functions, the possible Pauli
principle restriction has to be considered. In the
diquark-antidiquark configuration, we have
\begin{eqnarray}
\phi_1\chi_1&=&|[(Q_1Q_2)_1^6(\bar{Q}_3\bar{q}_4)_1^{\bar{6}}\rangle_2\delta_{12},\nonumber\\
\phi_1\chi_2&=&|(Q_1Q_2)_1^6(\bar{Q}_3\bar{q}_4)_1^{\bar{6}}\rangle_1\delta_{12}, \nonumber\\
\phi_1\chi_3&=&|(Q_1Q_2)_1^6(\bar{Q}_3\bar{q}_4)_1^{\bar{6}}\rangle_0\delta_{12}, \nonumber\\
\phi_1\chi_4&=&|(Q_1Q_2)_1^6(\bar{Q}_3\bar{q}_4)_0^{\bar{6}}\rangle_1\delta_{12},\nonumber\\
\phi_1\chi_5&=&|(Q_1Q_2)_0^6(\bar{Q}_3\bar{q}_4)_1^{\bar{6}}\rangle_1, \nonumber\\
\phi_1\chi_6&=&|(Q_1Q_2)_0^6(\bar{Q}_3\bar{q}_4)_0^{\bar{6}}\rangle_0,\\
\phi_2\chi_1&=&|(Q_1Q_2)_1^{\bar{3}}(\bar{Q}_3\bar{q}_4)^{3}_1\rangle_2,\nonumber\\
\phi_2\chi_2&=&|(Q_1Q_2)_1^{\bar{3}}(\bar{Q}_3\bar{q}_4)^{3}_1\rangle_1,\nonumber\\
\phi_2\chi_3&=&|(Q_1Q_2)_1^{\bar{3}}(\bar{Q}_3\bar{q}_4)^{3}_1\rangle_0,\nonumber\\
\phi_2\chi_4&=&|(Q_1Q_2)_1^{\bar{3}}(\bar{Q}_3\bar{q}_4)^{3}_0\rangle_1,\nonumber\\
\phi_2\chi_5&=&|(Q_1Q_2)_0^{\bar{3}}(\bar{Q}_3\bar{q}_4)^{3}_1\rangle_1\delta_{12},\nonumber\\
\phi_2\chi_6&=&|(Q_1Q_2)_0^{\bar{3}}(\bar{Q}_3\bar{q}_4)^{3}_0\rangle_0\delta_{12},
\end{eqnarray}
where $\delta_{12}=0$ if $Q_1$ and $Q_2$ are identical quarks, and
$\delta_{12}=1$ for the other cases. Similarly, the
triquark-antiquark base vectors in the $color\otimes spin$ space are
\begin{eqnarray}
\phi_1\zeta_1&=&|[(Q_1Q_2)_1\bar{Q}_3]^{3_{MS}}_\frac{3}{2}\bar{q}_4\rangle_2\delta_{12},\nonumber\\
\phi_1\zeta_2&=&|[(Q_1Q_2)_1\bar{Q}_3]^{3_{MS}}_\frac{3}{2}\bar{q}_4\rangle_1\delta_{12},\nonumber\\
\phi_1\zeta_3&=&|[(Q_1Q_2)_1\bar{Q}_3]^{3_{MS}}_\frac{1}{2}\bar{q}_4\rangle_1\delta_{12},\nonumber\\
\phi_1\zeta_5&=&|[(Q_1Q_2)_0\bar{Q}_3]^{3_{MS}}_\frac{1}{2}\bar{q}_4\rangle_1,\nonumber\\
\phi_1\zeta_4&=&|[(Q_1Q_2)_1\bar{Q}_3]^{3_{MS}}_\frac{1}{2}\bar{q}_4\rangle_0\delta_{12},\nonumber\\
\phi_1\zeta_6&=&|[(Q_1Q_2)_0\bar{Q}_3]^{3_{MS}}_\frac{1}{2}\bar{q}_4\rangle_0,\\
\phi_2\zeta_1&=&|[(Q_1Q_2)_1\bar{Q}_3]^{3_{MA}}_\frac{3}{2}\bar{q}_4\rangle_2,\nonumber\\
\phi_2\zeta_2&=&|[(Q_1Q_2)_1\bar{Q}_3]^{3_{MA}}_\frac{3}{2}\bar{q}_4\rangle_1,\nonumber\\
\phi_2\zeta_3&=&|[(Q_1Q_2)_1\bar{Q}_3]^{3_{MA}}_\frac{1}{2}\bar{q}_4\rangle_1,\nonumber\\
\phi_2\zeta_5&=&|[(Q_1Q_2)_0\bar{Q}_3]^{3_{MA}}_\frac{1}{2}\bar{q}_4\rangle_1\delta_{12},\nonumber\\
\phi_2\zeta_4&=&|[(Q_1Q_2)_1\bar{Q}_3]^{3_{MA}}_\frac{1}{2}\bar{q}_4\rangle_0,\nonumber\\
\phi_2\zeta_6&=&|[(Q_1Q_2)_0\bar{Q}_3]^{3_{MA}}_\frac{1}{2}\bar{q}_4\rangle_0\delta_{12}.
\end{eqnarray}

To exhaust all possible configurations of the $QQ\bar{Q}\bar{q}$
system, one replaces each $Q$ by either $c$ or $b$ quark. The cases
we need to study are: $bb\bar{b}\bar{q}$, $bb\bar{c}\bar{q}$,
$bc\bar{b}\bar{q}$, $bc\bar{c}\bar{q}$, $cc\bar{c}\bar{q}$, and
$cc\bar{b}\bar{q}$ ($q=u,d,s$). They can be divided into two
classes: (1) $bb\bar{b}\bar{q}$, $bb\bar{c}\bar{q}$,
$cc\bar{c}\bar{q}$ and $cc\bar{b}\bar{q}$; and (2)
$bc\bar{b}\bar{q}$ and $bc\bar{c}\bar{q}$. Because of the Pauli
principle for the first class, $\delta_{12}=0$ has to be adopted and
thus there are 6 independent $color\otimes spin$ bases. The second
class is not constrained by the Pauli principle and there are twelve
independent bases.

\subsection{The $bb\bar{b}\bar{q}$, $bb\bar{c}\bar{q}$, $cc\bar{c}\bar{q}$ and $cc\bar{b}\bar{q}$ systems}

The quantum numbers of these systems are $I(J^P)=\frac{1}{2}(2^+)$,
$\frac{1}{2}(1^+)$, or $\frac{1}{2}(0^+)$ when $q=u,d$. The isospin
is 0 if $q=s$. To write the CMI (color-magnetic interaction)
matrices in a convenient form, we define the combinations of the
effective couplings: $\mu=3C_{12}-C_{34}$, $\nu=C_{12}-3C_{34}$,
$\alpha=C_{12}+C_{34}$, $\beta=C_{13}+C_{14}$,
$\gamma=C_{23}+C_{24}$, $\eta=C_{13}-C_{14}$, $\tau=C_{23}-C_{24}$,
$\beta'=C_{13}+C_{23}$, $\gamma'=C_{14}+C_{24}$,
$\eta'=C_{13}-C_{23}$, and $\tau'=C_{14}-C_{24}$.

In the case $J=2$, there is only one state:
$\phi_2\chi_1=\psi_2\zeta_1$. The average of the CMI is $\langle
H_{CM}\rangle=\frac{8}{3}(\alpha+\beta)$ in both the
diquark-antidiquark and triquark-antiquark configurations. For a
state with given quark content, one replaces the $C_{ij}$ with the
appropriate number. For example, for the $bb\bar{b}\bar{q}$ system,
$C_{12}=C_{bb}$, $C_{13}=C_{23}=C_{b\bar{b}}$, and so on.

With the diquark-antidiquark base $(\phi_2\chi_3,\phi_1\chi_6)^{T}$,
the CMI matrix in the $J=0$ case is
\begin{eqnarray}
\langle H_{CM}\rangle=\left(\begin{array}{cc}
\frac{8}{3}(\alpha-2\beta)&4\sqrt{6}\beta\\
&4\alpha\\
\end{array}\right).
\end{eqnarray}
Since
$(\phi_2\zeta_4,\phi_1\zeta_6)^T=(\phi_2\chi_3,\phi_1\chi_6)^{T}$,
one also gets this matrix in the triquark-antiquark configuration.

In the $J=1$ case, one has different CMI matrix elements for the two
configurations. The obtained matrix is
\begin{eqnarray}
\langle H_{CM}\rangle= \left(\begin{array}{ccc}
\frac{8}{3}(\alpha-\beta) &\frac{8\sqrt{2}}{3}\eta&8\eta \\
&\frac{8}{3}\nu&-4\sqrt{2}\beta\\
&&\frac{4}{3}\mu
\end{array}\right)
\end{eqnarray}
with the diquark-antidiqurk base vector $(\phi_2\chi_2,
\phi_2\chi_4, \phi_1\chi_5)^T$. In the triquark-antiquark
configuration, one gets
\begin{eqnarray}
&&\langle H_{CM}\rangle=\frac49\times \nonumber\\
&&\left(\begin{array}{ccc}
(\mu+3\nu+3\beta'-5\gamma') &\sqrt{2}(\mu-3\nu-2\gamma')&-6\sqrt{3}\gamma' \\
&2(\mu-3\beta'+\gamma')&\frac{9}{\sqrt{6}}(3\beta'-\gamma')\\
&&3\mu
\end{array}\right)\nonumber\\
\end{eqnarray}
with the base vector $(\phi_{2}\zeta_{2}, \phi_{2}\zeta_{3},
\phi_{1}\zeta_{5} )^{T}$. The last matrix elements in the two
configurations are the same because
$\phi_{1}\zeta_{5}=\phi_1\chi_5$.

\subsection{The $bc\bar{b}\bar{q}$ and $bc\bar{c}\bar{q}$ systems}

The wave functions are not constrained by the Pauli principle and we
have $\delta_{12}=1$. When the total spin of such systems is $J=2$,
the allowed base states are $\phi_{1}\chi_{1}$ and
$\phi_{2}\chi_{1}$ in the diquark-antidiquark configuration. Then
the CMI matrix is \begin{eqnarray} \langle H_{CM}\rangle=
\left(\begin{array}{cc}
\frac{4}{3}(2\alpha+\beta+\gamma)&-2\sqrt{2}(\eta-\tau)\\
&\frac{2}{3}(-2\alpha+5\beta+5\gamma)
\end{array}\right).
\end{eqnarray}

We use the base vector $(\phi_{2}\chi_{3}, \phi_{2}\chi_{6},
\phi_{1}\chi_{3}, \phi_{1}\chi_{6} )^T$  for the CMI matrix in the
diquark-antidiquark configuration when the total spin is 0 and get

\begin{eqnarray}
&&\langle H_{CM}\rangle= \nonumber\\
&&\left(\begin{array}{cccc}
\frac{8}{3}(\alpha-\beta-\gamma)&\frac{4}{\sqrt{3}}(\tau-\eta)&4\sqrt{2}(\eta-\tau)&2\sqrt{6}(\beta+\gamma)\\
&-8\alpha&2\sqrt{6}(\beta+\gamma)&0\\
&&\frac{4}{3}\left(\begin{array}{c}-\alpha-\\5\beta-5\gamma\end{array}\right)&\frac{10}{\sqrt{3}}(\tau-\eta)\\
&&&4\alpha
\end{array}\right).\nonumber\\
\end{eqnarray}

In the case of $J=1$, we use $(\phi_{1}\chi_{2}$,
$\phi_{1}\chi_{4}$, $\phi_{1}\chi_{5}$, $\phi_{2}\chi_{2}$,
$\phi_{2}\chi_{4}$, $\phi_{2}\chi_{5} )^{T}$ as the base vector for
CMI in the diquark-antidiquark configuration. The obtained CMI
matrix are given in eq. (\ref{di}) below.

\begin{figure*}
\normalsize
\begin{eqnarray}
\label{di} \langle H_{CM}\rangle= \left(\begin{array}{cccccc}
-\frac{2}{3}(2\alpha+5\beta+5\gamma)&\frac{10\sqrt{2}}{3}(\eta+\tau)
&\frac{-10\sqrt{2}}{3}(\beta-\gamma)&2\sqrt{2}(\eta-\tau)&-4(\beta-\gamma)&4(\eta+\tau)\\
&-\frac{4}{3}\nu&\frac{10}{3}(\eta-\tau)&-4(\beta-\gamma)&0&-2\sqrt{2}(\beta+\gamma)\\
&&\frac{4}{3}\mu&4(\eta+\tau)&-2\sqrt{2}(\beta+\gamma)&0\\
&&&\frac{4}{3}(2\alpha-\beta-\gamma)&\frac{4\sqrt{2}}{3}(\eta+\tau)&\frac{4\sqrt{2}}{3}(-\beta+\gamma)\\
&&&&\frac{8}{3}\nu&\frac{4}{3}(\eta-\tau)\\
&&&&&-\frac{8}{3}\mu
\end{array}\right).
\end{eqnarray}
\end{figure*}

From
$(\phi_{1}\zeta_{1},\phi_{2}\zeta_{1})^T=(\phi_{1}\chi_{1},\phi_{2}\chi_{1})^T$
and $(\phi_{1}\zeta_4$, $\phi_{1}\zeta_{6}$, $\phi_{2}\zeta_4$,
$\phi_{2}\zeta_6)^T=(\phi_{1}\chi_{3}, \phi_{1}\chi_{6},
\phi_{2}\chi_{3}, \phi_{2}\chi_{6} )^T$, it is easy to understand
that the CMI matrices in the triquark-antiquark configuration are
the same as the above ones for the tensor ($J=2$) and scalar ($J=0$)
systems, respectively. The difference only occurs in the case of
$J=1$. If we choose the base vector $(\phi_{1}\zeta_{2}$,
$\phi_{1}\zeta_{3}$, $\phi_{1}\zeta_{5}$, $\phi_{2}\zeta_{2}$,
$\phi_{2}\zeta_{3}$, $\phi_{2}\zeta_{5} )^{T}$, the obtained CMI
matrix in the triquark-antiquark configuration is shown in eq.
(\ref{tri}) below.
\begin{figure*}
\normalsize
\begin{eqnarray}
\label{tri} \langle H_{CM}\rangle=\frac29
\left(\begin{array}{cccccc}
\left(\begin{array}{c}-\mu+15\beta'\\-3\nu-25\gamma'\end{array}\right)
&\sqrt{2}\left(\begin{array}{c}3\nu-\mu\\-10\gamma'\end{array}\right)&-10\sqrt{6}\tau'&-3\sqrt{2}(3\eta'+5\tau')
&-12\tau'&-12\sqrt{3}\gamma'\\
&2\left(\begin{array}{c}5\gamma'-\mu\\-15\beta'\end{array}\right)
&-5\sqrt{3}(3\eta'+\tau')
&-12\tau'&6\sqrt{2}(3\eta'+\tau')&3\sqrt{6}(3\beta'-\gamma')\\
&&6\mu&-12\sqrt{3}\gamma'&3\sqrt{6}(3\beta'-\gamma')&0\\
&&&2\left(\begin{array}{c}\mu+3\nu+\\3\beta'-5\gamma'\end{array}\right)
&2\sqrt{2}\left(\begin{array}{c}\mu-3\nu\\-2\gamma'\end{array}\right)
&-4\sqrt{6}\tau'\\
&&&&4\left(\begin{array}{c}\mu+\gamma'\\-3\beta'\end{array}\right)&-2\sqrt{3}(3\eta'+\tau')\\
&&&&&-12\mu
\end{array}\right).
\end{eqnarray}
\end{figure*}

\section{Numerical results}\label{sec3}

\subsection{Parameters}

The parameters $C_{Qq}$ ($Q=c, b$, $q=n,s$ with $n=u,d$) can be
extracted from the masses of charmed and bottom baryons while the
parameters $C_{Q\bar{Q}}$ ($C_{Q\bar{q}}$) are determined from the
heavy quarkonium ($D$ and $B$) mesons. We list the derived
parameters in Tab. \ref{parameter}. For $C_{b\bar{c}}$, $C_{bc}$,
$C_{bb}$ and $C_{cc}$, since the relevant baryons are not observed
in experiments, we use an estimation $C_{b\bar{c}}=3.3$ MeV from a
quark model calculation \cite{Godfrey:1985xj} and use the
approximation $C_{cc}=C_{c\bar{c}}$, $C_{bc}=C_{b\bar{c}}$,
$C_{cc}=C_{c\bar{c}}$ and $C_{bb}=C_{b\bar{b}}$.

\begin{table*}[!h]
\caption{The effective coupling parameters in units of
MeV.}\label{parameter} \centering
\begin{tabular}{cccccc}
\hline \hline
Hadron&$\langle H_{CM}\rangle$&Hadron&$\langle H_{CM}\rangle$&Parameter\\
\hline
$N$&$-8C_{nn}$&$\Delta$&$8C_{nn}$&$C_{nn}=18.4$\\
$\Sigma$&$\frac{8}{3}C_{nn}-\frac{32}{3}C_{n s}$&$\Sigma^*$&$\frac{8}{3}C_{nn}+\frac{16}{3}C_{n s}$&$C_{n s}=12.4$\\
$\Xi^0$&$\frac{8}{3}(C_{ss}-4C_{n s})$&$\Xi^{*0}$&$\frac{8}{3}(C_{ss}+C_{n s})$&\\
$\Omega$&8$C_{ss}$&&&$C_{ss}=6.4$\\
$D$&$-16C_{c\bar{n}}$&$D^{*}$&$\frac{16}{3}C_{c\bar{n}}$&$C_{c\bar{n}}=6.7$\\
$D_s$&$-16C_{c\bar{s}}$&$D_{s}^{*}$&$\frac{16}{3}C_{c\bar{s}}$&$C_{c\bar{s}}=6.7$\\
$B$&$-16C_{b\bar{n}}$&$B^{*}$&$\frac{16}{3}C_{b\bar{n}}$&$C_{b\bar{n}}=2.1$\\
$B_s$&$-16C_{b\bar{s}}$&$B^{*}$&$\frac{16}{3}C_{b\bar{s}}$&$C_{b\bar{s}}=2.3$\\
$\eta_{c}$&$-16C_{c\bar{c}}$&$J/\psi$&$\frac{16}{3}C_{c\bar{c}}$&$C_{c\bar{c}}=5.3$\\
$\eta_{b}$&$-16C_{b\bar{b}}$&$\Upsilon$&$\frac{16}{3}C_{b\bar{b}}$&$C_{b\bar{b}}=2.9$\\
$\Sigma_{c}$&$\frac{8}{3}C_{nn}-\frac{32}{3}C_{cn}$&$\Sigma_{c}^*$&$\frac{8}{3}C_{nn}+\frac{16}{3}C_{cn}$&$C_{cn}=4.0$\\
$\Xi'_{c}$&$\frac{8}{3}C_{n s}-\frac{16}{3}C_{cn}-\frac{16}{3}C_{cs}$&$\Xi_{c}^*$&$\frac{8}{3}C_{n s}+\frac{8}{3}C_{cn}+\frac{8}{3}C_{cs}$&$C_{cs}=4.6$\\
$\Sigma_{b}$&$\frac{8}{3}C_{nn}-\frac{32}{3}C_{bn}$&$\Sigma_{b}^{*}$&$\frac{8}{3}C_{nn}+\frac{16}{3}C_{bn}$&$C_{bn}=1.3$\\
$\Xi'_{b}$&$\frac{8}{3}C_{n
s}-\frac{16}{3}C_{bn}-\frac{16}{3}C_{bs}$&$\Xi_b^*$&
$\frac{8}{3}C_{n s}+\frac{8}{3}C_{bn}+\frac{8}{3}C_{bs}$&$C_{bs}=1.2$\\
\hline
\end{tabular}
\end{table*}

The present study is not a dynamical calculation and we use two
schemes to estimate roughly the mass of the $QQ\bar{Q}\bar{q}$
systems. In the first scheme, we use the effective quark masses
$m_{c}^{eff}=1724.8$ MeV, $m_{b}^{eff}=5052.9$ MeV,
$m_{n}^{eff}=361.8$ MeV and $m_{s}^{eff}=540.4$ MeV as inputs. These
masses are extracted from the known baryons. We have shown in Ref.
\cite{Wu:2016gas} that these values lead to overestimated meson
masses and give an upper limit for the ground state tetraquarks. In
the second scheme, we determine the tetraquark masses by comparing
to the threshold of a two-meson system. We will mainly focus on the
results in the second scheme. The meson masses we will use in this
work are \cite{Agashe:2014kda}: $m_{\Upsilon}=9460.3$ MeV,
$m_{\eta_b}=9398.0$ MeV, $m_{J/\psi}=3096.9$ MeV,
$m_{\eta_c}=2983.6$ MeV, $m_{D}=1864.8$ MeV, $m_{D^*}=2007.0$ MeV,
$m_{D_{s}}=1968.3$ MeV, $m_{D_{s}^*}=2112.1$ MeV, $m_{B}=5279.4$
MeV, $m_{B^*}=5325.2$ MeV, $m_{B_s}=5366.8$ MeV, $m_{B_s^*}=5415.4$
MeV, and $m_{B_c}=6275.6$ MeV.

\subsection{The $bb\bar{b}\bar{q}$, $bb\bar{c}\bar{q}$, $cc\bar{c}\bar{q}$ and $cc\bar{b}\bar{q}$ systems in diquark-antidiquark configuration}

By substituting the parameters into the CMI matrices in the previous
section and diagonalizing the matrices, we obtain the eigenvalues of
the CMI and the tetraquark masses with $M=\sum_{i}m_i+\langle
H_{CM}\rangle$. In the second scheme, we use the formula
$M=M_{ref}-\langle H_{CM}\rangle_{ref}+\langle H_{CM}\rangle$, where $ref$ means a reference two-meson system and $M_{ref}$ is its threshold.

\begin{table*}[h!]
\caption{Results for the $cc\bar{c}\bar{q}$ systems in units of MeV.
The masses in the fifth column are calculated with the effective
quark masses, which are theoretical upper limits. The last column
lists masses estimated from the $J/\psi D$ ($J/\psi D_s$)
threshold.}\label{result:cccq} \centering
\begin{tabular}{c|ccccccc}\hline
\multicolumn{6}{c}{$cc\bar{c}\bar{n}$ system} \\\hline\hline $J^{P}$
& $\langle H_{CM} \rangle$ &Eigenvalue &Eigenvector &Mass&$(J/\psi
D)$\\\hline
$2^{+}$ &56.8&56.8&1&5593.0&5097.4\\
$1^{+}$ &$\left(\begin{array}{ccc}-7.2&-5.3&-11.2\\-5.3&-17.9&-67.9\\-11.2&-67.9&15.9\end{array}\right)$&$\left(\begin{array}{c}-72.8\\69.4\\-5.7\end{array}\right)$&$\left[\begin{array}{ccc}\{-0.17,-0.77,-0.61\}\\\{-0.07,-0.61,0.79\}\\\{0.98,-0.18,-0.04\}\end{array}\right]$&$\left(\begin{array}{c}5463.4\\5605.6\\5530.5\end{array}\right)$&$\left(\begin{array}{c}4967.8\\5110.0\\5034.9\end{array}\right)$\\
$0^{+}$ &$\left(\begin{array}{cc}-39.2&117.6\\117.6&37.2\end{array}\right)$&$\left(\begin{array}{c}-124.6\\122.6\end{array}\right)$&$\left[\begin{array}{cc}\{-0.81,0.59\}\\\{-0.59,-0.81\}\end{array}\right]$&$\left(\begin{array}{c}5411.6\\5658.8\end{array}\right)$&$\left(\begin{array}{c}4916.0\\5163.3\end{array}\right)$\\
\hline\hline\multicolumn{6}{c}{$cc\bar{c}\bar{s}$ system}
\\\hline\hline $J^{P}$ & $\langle H_{CM} \rangle$ &Eigenvalue
&Eigenvector &Mass&$(J/\psi D_{s})$\\\hline
$2^{+}$ &58.4&58.4&1&5773.2&5202.5\\
$1^{+}$ &$\left(\begin{array}{ccc}-5.6&-5.3&-11.2\\-5.3&-22.7&-67.9\\-11.2&-67.9&15.1\end{array}\right)$&$\left(\begin{array}{c}-76.0\\67.1\\-4.3\end{array}\right)$&$\left[\begin{array}{ccc}\{-0.15,-0.78,-0.60\}\\\{-0.08,-0.60,0.80\}\\\{0.98,-0.17,-0.03\}\end{array}\right]$&$\left(\begin{array}{c}5638.8\\5781.9\\5710.5\end{array}\right)$&$\left(\begin{array}{c}5068.2\\5211.2\\5139.8\end{array}\right)$\\
$0^{+}$ &$\left(\begin{array}{cc}-37.6&117.6\\117.6&39.6\end{array}\right)$&$\left(\begin{array}{c}124.7\\-122.7\end{array}\right)$&$\left[\begin{array}{cc}\{0.59,0.81\}\\\{-0.81,0.59\}\end{array}\right]$&$\left(\begin{array}{c}5839.5\\5592.1\end{array}\right)$&$\left(\begin{array}{c}5268.9\\5021.4\end{array}\right)$\\
\hline
\end{tabular}
\end{table*}

\begin{figure*}[!h]\centering
\begin{tabular}{lcr}
\resizebox{0.45\textwidth}{!}{\includegraphics{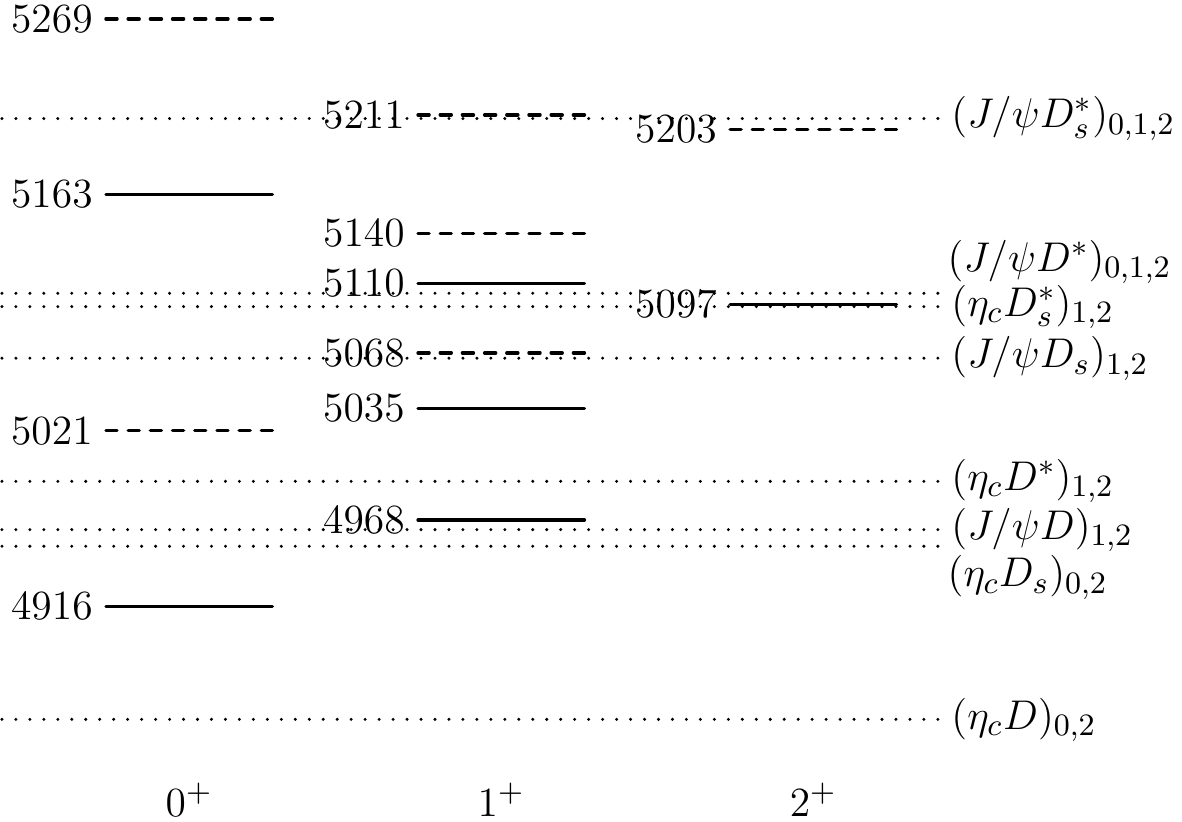}}&\qquad&
\resizebox{0.45\textwidth}{!}{\includegraphics{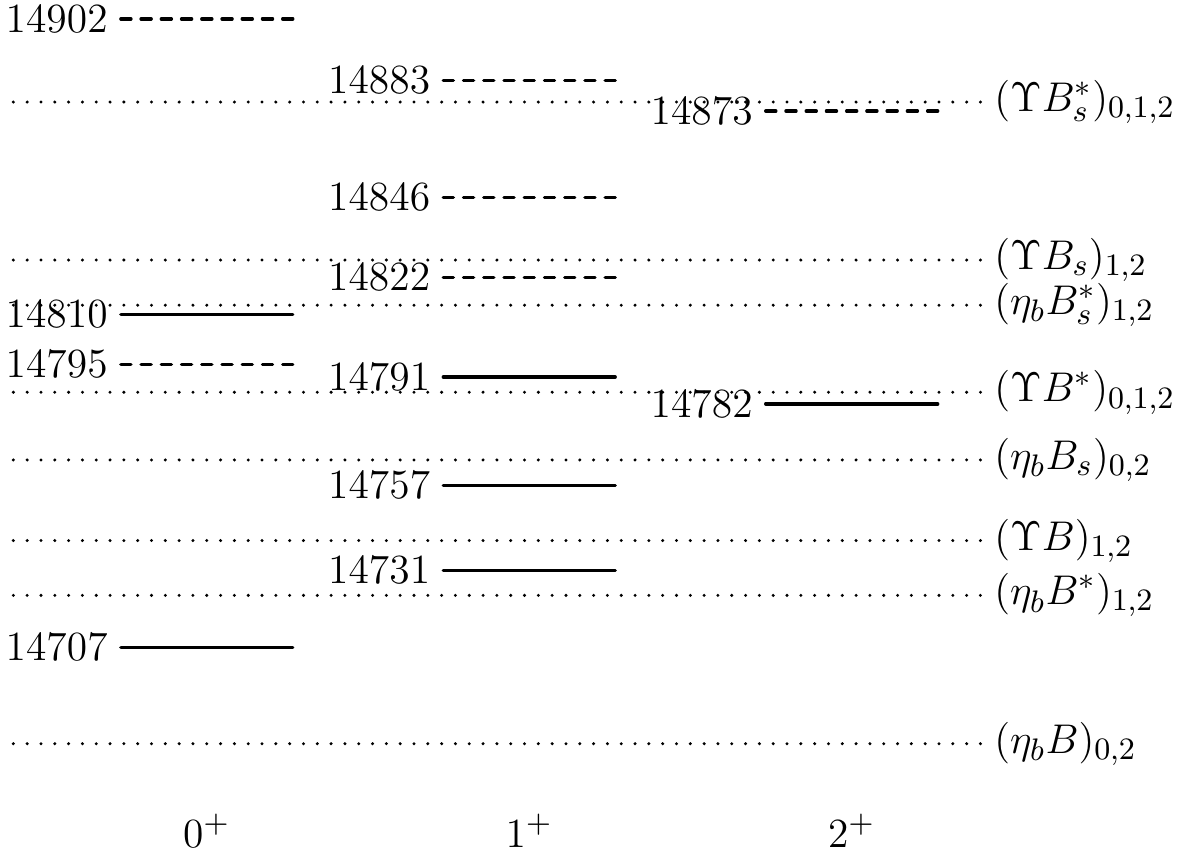}}
\end{tabular}
\caption{Proposed $cc\bar{c}\bar{q}$ (left) and $bb\bar{b}\bar{q}$
(right) tetraquark states. The solid (dashed) line corresponds to
the case $q=u,d$ ($q=s$). The dotted line indicates various
meson-meson thresholds. When a number in the subscript of a
meson-meson state is equal to the spin of an initial state, the
decay for the initial state into that meson-meson channel through
$S$- or $D$-wave is allowed. The masses are given in units of
MeV.}\label{QQQ-q}
\end{figure*}

We present the CMI matrices, eigenvalues, eigenvectors and the
estimated masses for the $cc\bar{c}\bar{n}$ ($n=u,d$) and the
$cc\bar{c}\bar{s}$ systems in Table \ref{result:cccq}. If these
states really exist, probably the masses are slightly above the
values in the last column. The reason is that the mass estimated
with a color-spin interaction seems underestimated and a correction
from the additional kinetic energy is probably needed
\cite{Wu:2016gas,Park:2015nha}. The mass splitting between the
$cc\bar{c}\bar{n}$ tetraquarks with different spins is at most 250
MeV. The maximum splitting for the $cc\bar{c}\bar{s}$ tetraquarks is
similar. We plot the relative positions of these states in Fig.
\ref{QQQ-q}, where the masses in the threshold approach are adopted.

The quantum numbers of these $cc\bar{c}\bar{n}$ ($cc\bar{c}\bar{s}$)
tetraquarks are the same as those of $D_0^*$, $D_1$, and $D_2^*$
($D_{s0}^*$, $D_{s1}$, and $D_{s2}^*$). The orbital or radial
excitation cannot induce a state 2500 MeV higher than the ground
state. Therefore, once the predicted states could be observed, it is
easy to identify them as $D$- or $D_s^+$-like mesons with an excited
charm-anticharm pair. The argument is similar to the work in
predicting the hidden charm pentaquarks \cite{Wu:2010jy}. Compared
with the pentaquark case, the binding force for the present system
is dominantly provided by the gluon exchange interaction. The
contribution from the meson exchange is highly suppressed and the
interaction between a charmonium and a heavy-light meson is not
strong. As a result, a molecule or cusp interpretation is not
favored once the resonance structure is observed. In this sense, one
may observe a genuine tetraquark.

\begin{table*}[h!]
\caption{Results for the $bb\bar{b}\bar{q}$ systems in units of MeV.
The masses in the fifth column are calculated with the effective
quark masses and are theoretical upper limits. The last column lists
masses estimated from the $\Upsilon B$ ($\Upsilon B_s$) threshold.
}\label{result:bbbq} \centering
\begin{tabular}{c|ccccccc}\hline
\multicolumn{6}{c}{$bb\bar{b}\bar{n}$ system} \\\hline\hline $J^{P}$
& $\langle H_{CM} \rangle$ &Eigenvalue &Eigenvector &Mass&$(\Upsilon
B)$\\\hline
$2^{+}$ &24.5&24.5&1&15545.0&14782.4\\
$1^{+}$ &$\left(\begin{array}{ccc}-2.1&3.0&6.4\\3.0&-2.7&-28.3\\6.4&-28.3&9.9\end{array}\right)$&$\left(\begin{array}{c}32.9\\-27.0\\-0.8\end{array}\right)$&$\left[\begin{array}{ccc}\{0.09,-0.62,0.78\}\\\{0.25,-0.75,-0.62\}\\\{0.96,0.25,0.09\}\end{array}\right]$&$\left(\begin{array}{c}15553.4\\15493.5\\15519.7\end{array}\right)$&$\left(\begin{array}{c}14790.7\\14730.8\\14757.1\end{array}\right)$\\
$0^{+}$ &$\left(\begin{array}{cc}-15.5&49.0\\49.0&16.8\end{array}\right)$&$\left(\begin{array}{c}52.2\\-50.9\end{array}\right)$&$\left[\begin{array}{cc}\{0.59,0.81\}\\\{-0.81,0.59\}\end{array}\right]$&$\left(\begin{array}{c}15572.7\\15469.6\end{array}\right)$&$\left(\begin{array}{c}14810.1\\14706.9\end{array}\right)$\\
\hline\hline\multicolumn{6}{c}{$bb\bar{b}\bar{s}$ system}
\\\hline\hline $J^{P}$ & $\langle H_{CM} \rangle$ &Eigenvalue
&Eigenvector &Mass&$(\Upsilon B_{s})$\\\hline
$2^{+}$ &24.8&24.8&1&15723.9&14873.2\\
$1^{+}$ &$\left(\begin{array}{ccc}-2.9&2.3&4.8\\2.3&-1.9&-29.4\\4.8&-29.4&10.0\end{array}\right)$&$\left(\begin{array}{c}34.2\\-26.9\\-2.1\end{array}\right)$&$\left[\begin{array}{ccc}\{0.06,-0.63,0.78\}\\\{0.20,-0.75,-0.63\}\\\{0.98,0.19,0.08\}\end{array}\right]$&$\left(\begin{array}{c}15733.3\\15672.2\\15697.0\end{array}\right)$&$\left(\begin{array}{c}14882.6\\14821.5\\14846.3\end{array}\right)$\\
$0^{+}$ &$\left(\begin{array}{cc}-16.8&50.9\\50.9&16.4\end{array}\right)$&$\left(\begin{array}{c}-53.8\\53.4\end{array}\right)$&$\left[\begin{array}{cc}\{-0.81,0.59\}\\\{-0.59,-0.81\}\end{array}\right]$&$\left(\begin{array}{c}15645.3\\15752.5\end{array}\right)$&$\left(\begin{array}{c}14794.6\\14901.8\end{array}\right)$\\
\hline
\end{tabular}
\end{table*}

\begin{table*}[!h]
\caption{Results for the $cc\bar{b}\bar{q}$ systems in units of MeV.
The masses in the fifth column are calculated with the effective
quark masses and are theoretical upper limits. The last column lists
masses estimated from the $B_{c} D$ ($B_{c} D_s$)
threshold.}\label{result:ccbq} \centering
\begin{tabular}{c|ccccccc}\hline
\multicolumn{6}{c}{$cc\bar{b}\bar{n}$ system} \\\hline\hline $J^{P}$
& $\langle H_{CM} \rangle$ &Eigenvalue &Eigenvector &Mass&$(B_c
D)$\\\hline
$2^{+}$ &44.3&44.3&1&8908.6&8344.7\\
$1^{+}$ &$\left(\begin{array}{ccc}-9.1&-12.8&-27.2\\-12.8&3.7&-56.6\\-27.2&-56.6&19.5\end{array}\right)$&$\left(\begin{array}{c}70.7\\-60.6\\4.0\end{array}\right)$&$\left[\begin{array}{ccc}\{-0.16,-0.62,0.77\}\\\{-0.48,-0.63,-0.61\}\\\{0.86,-0.47,-0.19\}\end{array}\right]$&$\left(\begin{array}{c}8935.0\\8803.7\\8868.3\end{array}\right)$&$\left(\begin{array}{c}8371.1\\8239.8\\8304.4\end{array}\right)$\\
$0^{+}$ &$\left(\begin{array}{cc}-35.7&98.0\\98.0&26.4\end{array}\right)$&$\left(\begin{array}{c}-107.5\\98.1\end{array}\right)$&$\left[\begin{array}{cc}\{-0.81,0.59\}\\\{-0.59,-0.81\}\end{array}\right]$&$\left(\begin{array}{c}8756.8\\8962.4\end{array}\right)$&$\left(\begin{array}{c}8192.9\\8398.5\end{array}\right)$\\
\hline\hline\multicolumn{6}{c}{$cc\bar{b}\bar{s}$ system}
\\\hline\hline $J^{P}$ & $\langle H_{CM} \rangle$ &Eigenvalue
&Eigenvector &Mass&$(B_c D_{s})$\\\hline
$2^{+}$ &44.0&44.0&1&9086.9&8447.9\\
$1^{+}$ &$\left(\begin{array}{ccc}-9.3&-12.8&-27.2\\-12.8&4.5&-56.6\\-27.2&-56.6&19.6\end{array}\right)$&$\left(\begin{array}{c}71.1\\-60.3\\4.0\end{array}\right)$&$\left[\begin{array}{ccc}\{-0.16,-0.62,0.77\}\\\{-0.48,-0.63,-0.61\}\\\{0.86,-0.47,-0.20\}\end{array}\right]$&$\left(\begin{array}{c}9114.0\\8982.6\\9046.9\end{array}\right)$&$\left(\begin{array}{c}8475.0\\8343.6\\8407.9\end{array}\right)$\\
$0^{+}$ &$\left(\begin{array}{cc}-36.0&98.0\\98.0&26.0\end{array}\right)$&$\left(\begin{array}{c}-107.8\\97.8\end{array}\right)$&$\left[\begin{array}{cc}\{-0.81,0.59\}\\\{-0.59,-0.81\}\end{array}\right]$&$\left(\begin{array}{c}8935.1\\9140.7\end{array}\right)$&$\left(\begin{array}{c}8296.1\\8501.7\end{array}\right)$\\
\hline
\end{tabular}
\end{table*}

Let us take a look at the possible $S$-wave strong decay channels of
these compact tetraquark states from Fig. \ref{QQQ-q}. For the
scalar states, the dominant channels are possibly $\eta_c D$,
$\eta_cD_s$, $J/\psi D^*$, and $J/\psi D_s^*$. For the axial vector
states, possible decay channels are $J/\psi D$, $J/\psi D^*$,
$\eta_c D^*$, $J/\psi D_s$, $J/\psi D_s^*$, $\eta_c D_s^*$. For the
tensor states, the possible decay channels are $J/\psi D^*$ and
$J/\psi D_s^*$. Whether relevant channels are open or not depends on
the tetraquark mass and flavor conservation. For example, the tensor
meson with $M=5097$ (5203) is around the threshold of $J/\psi D^*$
($J/\psi D_s^*$) and the decay is marginal. However, it is still
possible to find a resonance slightly above the threshold if one
considers the limitation of the present estimation method. More
channels are possible if the $D$-wave decays are considered. In Fig.
\ref{QQQ-q}, we present several numbers in the subscript of the
meson-meson states. When a number is equal to the spin of an initial
state, the decay for the initial state into that meson-meson channel
through $S$- or $D$-wave is allowed.

By replacing the charm quark with the bottom quark, we get the
results for the $bb\bar{b}\bar{n}$ and $bb\bar{b}\bar{s}$ systems in
Table \ref{result:bbbq}. The rough position is also given in Fig.
\ref{QQQ-q} and similar analysis for the possible decay channels is
straightforward.

\begin{figure*}[!h]\centering
\begin{tabular}{lcr}
\resizebox{0.45\textwidth}{!}{\includegraphics{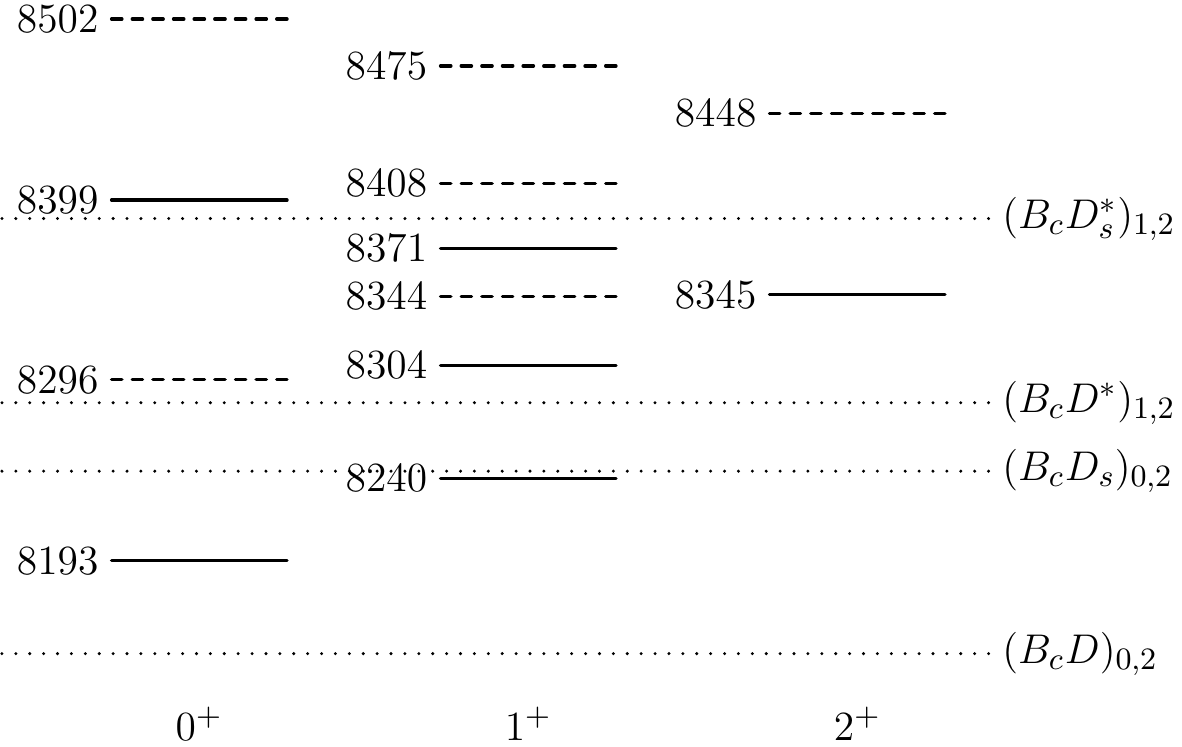}}&\qquad&
\resizebox{0.45\textwidth}{!}{\includegraphics{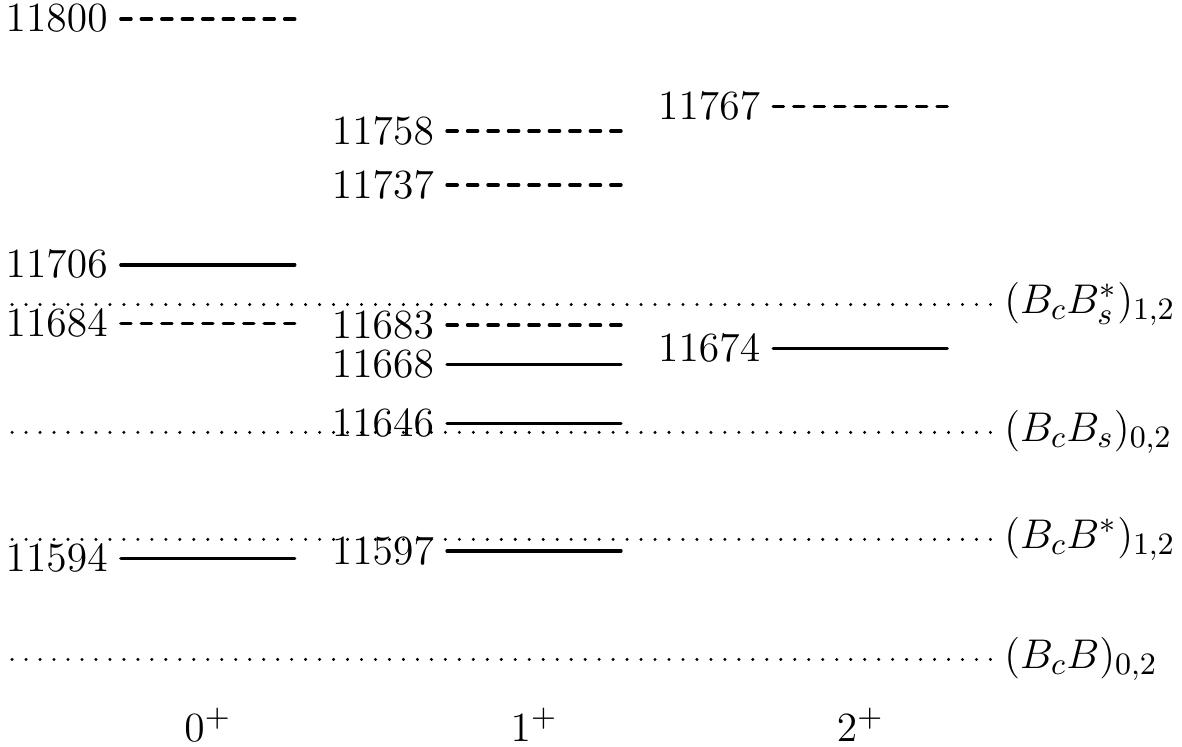}}
\end{tabular}
\caption{Proposed $cc\bar{b}\bar{q}$ (left) and $bb\bar{c}\bar{q}$
(right) tetraquark states. The solid (dashed) line corresponds to
the case $q=u,d$ ($q=s$). The dotted line indicates various
meson-meson thresholds. When a number in the subscript of a
meson-meson state is equal to the spin of an initial state, the
decay for the initial state into that meson-meson channel through
$S$- or $D$-wave is allowed. The masses are given in units of
MeV.}\label{QQQp-q}
\end{figure*}

Now we move on to the $cc\bar{b}\bar{q}$ and $bb\bar{c}\bar{q}$
systems. Such states are explicitly exotic. Up to now, the doubly
charmed baryon $\Xi_{cc}$ has not been confirmed since the first
announcement by the SELEX Collaboration \cite{Mattson:2002vu}. There
is no experimental result on the search for the proposed $T_{cc}$
($cc\bar{q}\bar{q}$) tetraquark. But their existence is not
excluded. If one replaces a light antiquark in the $T_{cc}$ with an
anti-bottom quark, one gets a currently discussed tetraquark system
$cc\bar{b}\bar{q}$. The exchange of bottom and charm results in
another exotic tetraquark system $bb\bar{c}\bar{q}$. We present the
results for the $cc\bar{b}\bar{q}$ and $bb\bar{c}\bar{q}$ systems in
Tabs. \ref{result:ccbq} and \ref{result:bbcq}, respectively. In Fig.
\ref{QQQp-q}, we display the rough positions of these possible
tetraquarks, where the masses in the threshold approach are adopted.
It is easy to judge their possible decays from the figure.

\begin{table*}[h!]
\caption{Results for the $bb\bar{c}\bar{q}$ systems in units of MeV.
The masses in the fifth column are calculated with the effective
quark masses and are theoretical upper limits. The last column lists
masses estimated from the $B_{c} B$ ($B_{c} B_s$)
threshold.}\label{result:bbcq} \centering
\begin{tabular}{c|ccccccc}\hline
\multicolumn{6}{c}{$bb\bar{c}\bar{n}$ system} \\\hline\hline $J^{P}$
& $\langle H_{CM} \rangle$ &Eigenvalue &Eigenvector &Mass&$(B_c
D)$\\\hline
$2^{+}$ &32.8&32.8&1&12225.2&11674.2\\
$1^{+}$ &$\left(\begin{array}{ccc}4.0&4.5&9.6\\4.5&-24.3&-30.5\\9.6&-30.5&6.3\end{array}\right)$&$\left(\begin{array}{c}-44.8\\26.7\\4.1\end{array}\right)$&$\left[\begin{array}{ccc}\{0.18,-0.83,-0.53\}\\\{0.26,-0.48,0.84\}\\\{0.95,0.29,-0.13\}\end{array}\right]$&$\left(\begin{array}{c}12147.6\\12219.1\\12196.5\end{array}\right)$&$\left(\begin{array}{c}11596.6\\11668.1\\11645.5\end{array}\right)$\\
$0^{+}$ &$\left(\begin{array}{cc}-10.4&52.9\\52.9&27.6\end{array}\right)$&$\left(\begin{array}{c}64.8\\-47.6\end{array}\right)$&$\left[\begin{array}{cc}\{0.58,0.82\}\\\{-0.82,0.58\}\end{array}\right]$&$\left(\begin{array}{c}12257.2\\12144.8\end{array}\right)$&$\left(\begin{array}{c}11706.2\\11593.8\end{array}\right)$\\
\hline\hline\multicolumn{6}{c}{$bb\bar{c}\bar{s}$ system}
\\\hline\hline $J^{P}$ & $\langle H_{CM} \rangle$ &Eigenvalue
&Eigenvector &Mass&$(B_c D_{s})$\\\hline
$2^{+}$ &34.9&34.9&1&12405.9&11766.9\\
$1^{+}$ &$\left(\begin{array}{ccc}5.1&3.8&8.0\\3.8&-29.1&-31.7\\8.0&-31.7&5.5\end{array}\right)$&$\left(\begin{array}{c}-48.9\\25.5\\4.8\end{array}\right)$&$\left[\begin{array}{ccc}\{0.14,-0.85,-0.51\}\\\{0.24,-0.47,0.85\}\\\{0.96,0.24,-0.14\}\end{array}\right]$&$\left(\begin{array}{c}12322.1\\12396.5\\12375.8\end{array}\right)$&$\left(\begin{array}{c}11683.1\\11757.5\\11736.8\end{array}\right)$\\
$0^{+}$ &$\left(\begin{array}{cc}-9.9&54.9\\54.9&30.0\end{array}\right)$&$\left(\begin{array}{c}68.4\\-48.3\end{array}\right)$&$\left[\begin{array}{cc}\{0.57,0.82\}\\\{-0.82,0.57\}\end{array}\right]$&$\left(\begin{array}{c}12439.4\\12322.7\end{array}\right)$&$\left(\begin{array}{c}11800.4\\11683.7\end{array}\right)$\\
\hline
\end{tabular}
\end{table*}

\subsection{The $bc\bar{b}\bar{q}$ and $bc\bar{c}\bar{q}$ systems in the diquark-antidiaquark configuration}

\begin{table*}[!h]\centering
\caption{Results for the $bc\bar{b}\bar{q}$ systems in units of MeV.
The masses in the fifth column are calculated with the effective
quark masses and are theoretical upper limits. The last two columns
list masses estimated from the $\Upsilon D$ ($\Upsilon D_{s}$) and
$B B_c$ ($B_s B_c$) thresholds, respectively.}\label{result:bcbq}
\scriptsize
\begin{tabular}{c|ccccccc}\hline
\multicolumn{6}{c}{$bc\bar{b}\bar{n}$ system} \\\hline\hline $J^{P}$
& $\langle H_{CM} \rangle$ &Eigenvalue &Eigenvector &Mass&$(\Upsilon
D)$&$(BB_c)$\\\hline
$2^{+}$ &$\left(\begin{array}{cc}32.3&-11.9\\-11.9&43.9\end{array}\right)$&$\left(\begin{array}{c}51.3\\24.8\end{array}\right)$&$\left[\begin{array}{cc}\{-0.53,0.85\}\\\{-0.85,-0.53\}\end{array}\right]$&$\left(\begin{array}{c}12243.7\\12217.2\end{array}\right)$&$\left(\begin{array}{c}11468.1\\11441.7\end{array}\right)$&$\left(\begin{array}{c}11692.7\\11666.2\end{array}\right)$\\
$1^{+}$ &$\left(\begin{array}{cccccc}-56.1&-12.3&23.6&11.9&20.0&-10.4\\-12.3&0.8&14.0&20.0&0.0&-42.4\\23.6&14.0&11.5&-10.4&-42.4&0.0\\11.9&20.0&-10.4&-7.7&-4.9&9.4\\20.0&0.0&-42.4&-4.9&-1.6&5.6\\-10.4&-42.4&0.0&9.4&5.6&-22.9\end{array}\right)$&$\left(\begin{array}{c}-95.6\\53.7\\-47.5\\31.0\\-15.6\\-2.2\end{array}\right)$&$\left[\begin{array}{cccccc}\{0.66,0.35,-0.35,-0.27,-0.34,0.36\}\\\{-0.03,-0.39,-0.69,-0.02,0.55,0.26\}\\\{-0.41,0.48,0.23,-0.18,0.29,0.66\}\\\{-0.03,0.70,-0.34,0.29,0.30,-0.46\}\\\{0.62,-0.05,0.48,0.15,0.60,0.03\}\\\{0.02,-0.03,-0.04,0.89,-0.23,0.40\}\end{array}\right]$&$\left(\begin{array}{c}12096.8\\12246.1\\12144.9\\12223.4\\12176.8\\12190.2\end{array}\right)$&$\left(\begin{array}{c}11321.3\\11470.5\\11369.4\\11447.8\\11401.3\\11414.7\end{array}\right)$&$\left(\begin{array}{c}11545.8\\11695.1\\11593.9\\11672.4\\11625.8\\11639.2\end{array}\right)$\\
$0^{+}$ &$\left(\begin{array}{cccc}-27.7&-9.7&23.8&73.5\\-9.7&-36.8&73.5&0.0\\23.8&73.5&-106.1&-24.2\\73.5&0.0&-24.2&18.4\end{array}\right)$&$\left(\begin{array}{c}-165.4\\73.3\\-70.5\\10.3\end{array}\right)$&$\left[\begin{array}{cccc}\{-0.29,-0.48,0.80,0.22\}\\\{0.58,-0.10,-0.07,0.81\}\\\{-0.74,0.31,-0.24,0.55\}\\\{0.18,0.82,0.55,0.02\}\end{array}\right]$&$\left(\begin{array}{c}12027.0\\12265.7\\12121.9\\12202.7\end{array}\right)$&$\left(\begin{array}{c}11251.4\\11490.2\\11346.4\\11427.1\end{array}\right)$&$\left(\begin{array}{c}11476.0\\11714.7\\11570.9\\11651.7\end{array}\right)$\\
\hline\hline\multicolumn{6}{c}{$bc\bar{b}\bar{s}$ system}
\\\hline\hline $J^{P}$ & $\langle H_{CM} \rangle$ &Eigenvalue
&Eigenvector &Mass&$(\Upsilon D_{s})$&$(B_sB_c)$\\\hline
$2^{+}$ &$\left(\begin{array}{cc}32.3&-11.3\\-11.3&44.7\end{array}\right)$&$\left(\begin{array}{c}51.4\\25.6\end{array}\right)$&$\left[\begin{array}{cc}\{-0.51,0.86\}\\\{-0.86,-0.51\}\end{array}\right]$&$\left(\begin{array}{c}12422.4\\12396.6\end{array}\right)$&$\left(\begin{array}{c}11571.7\\11545.9\end{array}\right)$&$\left(\begin{array}{c}11783.4\\11757.6\end{array}\right)$\\
$1^{+}$ &$\left(\begin{array}{cccccc}-56.7&-13.2&22.6&11.3&19.2&-11.2\\-13.2&0.4&13.3&19.2&0.0&-43.0\\22.6&13.3&11.6&-11.2&-43.0&0.0\\11.3&19.2&-11.2&-8.3&-5.3&9.1\\19.2&0.0&-43.0&-5.3&-0.8&5.3\\-11.2&-43.0&0.0&9.1&5.3&-23.2\end{array}\right)$&$\left(\begin{array}{c}-95.7\\54.0\\-47.0\\31.5\\-17.3\\-2.5\end{array}\right)$&$\left[\begin{array}{cccccc}\{0.66,0.36,-0.34,-0.27,-0.32,0.37\}\\\{-0.03,-0.37,-0.70,-0.00,0.56,0.25\}\\\{-0.41,0.47,0.24,-0.16,0.30,0.66\}\\\{-0.03,0.71,-0.32,0.28,0.28,-0.48\}\\\{0.62,-0.05,0.48,0.22,0.58,0.03\}\\\{-0.02,-0.02,-0.08,0.88,-0.27,0.37\}\end{array}\right]$&$\left(\begin{array}{c}12275.3\\12425.0\\12324.0\\12402.5\\12353.7\\12368.5\end{array}\right)$&$\left(\begin{array}{c}11424.6\\11574.4\\11473.3\\11551.8\\11503.1\\11517.8\end{array}\right)$&$\left(\begin{array}{c}11636.3\\11786.0\\11685.0\\11763.5\\11714.7\\11729.5\end{array}\right)$\\
$0^{+}$ &$\left(\begin{array}{cccc}-28.5&-9.2&22.6&74.5\\-9.2&-36.0&74.5&0.0\\22.6&74.5&-107.3&-23.1\\74.5&0.0&-23.1&18.0\end{array}\right)$&$\left(\begin{array}{c}-165.9\\73.6\\-72.9\\11.3\end{array}\right)$&$\left[\begin{array}{cccc}\{-0.28,-0.48,0.80,0.21\}\\\{0.58,-0.10,-0.07,0.81\}\\\{0.75,-0.29,0.24,-0.55\}\\\{0.17,0.82,0.54,0.03\}\end{array}\right]$&$\left(\begin{array}{c}12205.1\\12444.6\\12298.1\\12382.3\end{array}\right)$&$\left(\begin{array}{c}11354.4\\11594.0\\11447.5\\11531.6\end{array}\right)$&$\left(\begin{array}{c}11566.1\\11805.6\\11659.1\\11743.3\end{array}\right)$\\
\hline
\end{tabular}
\end{table*}

\begin{table*}[!h]
\caption{Results for the $bc\bar{c}\bar{q}$ systems in units of MeV.
The masses in the fifth column are calculated with the effective
quark masses and are theoretical upper limits. The last two columns
list masses estimated from the $B_{c} D$ ($B_{c} D_s$) and $BJ/\psi$
($B_sJ/\psi$) thresholds, respectively.}\label{result:bccq}
\centering\scriptsize
\begin{tabular}{c|ccccccc}\hline
\multicolumn{6}{c}{$bc\bar{c}\bar{n}$ system} \\\hline\hline $J^{P}$
& $\langle H_{CM} \rangle$ &Eigenvalue &Eigenvector
&Mass&$(B_cD)$&$(BJ/\psi)$\\\hline
$2^{+}$ &$\left(\begin{array}{cc}42.7&-7.4\\-7.4&48.3\end{array}\right)$&$\left(\begin{array}{c}53.3\\37.6\end{array}\right)$&$\left[\begin{array}{cc}\{-0.57,0.82\}\\\{-0.82,-0.57\}\end{array}\right]$&$\left(\begin{array}{c}8917.6\\8901.9\end{array}\right)$&$\left(\begin{array}{c}8353.7\\8338.0\end{array}\right)$&$\left(\begin{array}{c}8435.0\\8419.2\end{array}\right)$\\
$1^{+}$ &$\left(\begin{array}{cccccc}-67.7&-0.9&31.1&7.4&26.4&-0.8\\-0.9&11.6&8.7&26.4&0.0&-49.2\\31.1&8.7&7.9&-0.8&-49.2&0.0\\7.4&26.4&-0.8&-3.7&-0.4&12.4\\26.4&0.0&-49.2&-0.4&-23.2&3.5\\-0.8&-49.2&0.0&12.4&3.5&-15.7\end{array}\right)$&$\left(\begin{array}{c}-105.0\\-63.5\\56.7\\41.0\\-25.0\\4.9\end{array}\right)$&$\left[\begin{array}{cccccc}\{0.73,0.10,-0.43,-0.10,-0.50,0.09\}\\\{-0.14,0.58,0.04,-0.38,0.08,0.69\}\\\{0.06,0.70,0.42,0.22,-0.26,-0.45\}\\\{0.07,-0.39,0.71,-0.15,-0.50,0.28\}\\\{0.64,-0.01,0.37,-0.16,0.65,-0.04\}\\\{-0.13,-0.02,-0.05,-0.86,-0.08,-0.47\}\end{array}\right]$&$\left(\begin{array}{c}8759.3\\8800.8\\8921.0\\8905.3\\8839.3\\8869.2\end{array}\right)$&$\left(\begin{array}{c}8195.4\\8236.9\\8357.1\\8341.4\\8275.4\\8305.3\end{array}\right)$&$\left(\begin{array}{c}8276.6\\8318.2\\8438.3\\8422.6\\8356.7\\8386.5\end{array}\right)$\\
$0^{+}$ &$\left(\begin{array}{cccc}-26.9&-6.0&14.7&85.2\\-6.0&-58.4&85.2&0.0\\14.7&85.2&-125.7&-15.0\\85.2&0.0&-15.0&29.2\end{array}\right)$&$\left(\begin{array}{c}-187.7\\91.1\\-85.3\\-0.0\end{array}\right)$&$\left[\begin{array}{cccc}\{-0.16,-0.54,0.82,0.12\}\\\{0.58,-0.04,-0.03,0.81\}\\\{0.79,-0.18,0.11,-0.57\}\\\{0.10,0.82,0.57,-0.01\}\end{array}\right]$&$\left(\begin{array}{c}8676.6\\8955.4\\8779.0\\8864.3\end{array}\right)$&$\left(\begin{array}{c}8112.7\\8391.5\\8215.1\\8300.4\end{array}\right)$&$\left(\begin{array}{c}8193.9\\8472.8\\8296.3\\8381.6\end{array}\right)$\\
\hline\hline\multicolumn{6}{c}{$bc\bar{c}\bar{s}$ system}
\\\hline\hline $J^{P}$ & $\langle H_{CM} \rangle$ &Eigenvalue
&Eigenvector &Mass&$(B_cD_s)$&$(B_sJ/\psi)$\\\hline
$2^{+}$ &$\left(\begin{array}{cc}44.5&-6.8\\-6.8&48.1\end{array}\right)$&$\left(\begin{array}{c}53.4\\39.3\end{array}\right)$&$\left[\begin{array}{cc}\{-0.61,0.79\}\\\{-0.79,-0.61\}\end{array}\right]$&$\left(\begin{array}{c}9096.3\\9082.2\end{array}\right)$&$\left(\begin{array}{c}8457.3\\8443.2\end{array}\right)$&$\left(\begin{array}{c}8525.6\\8511.5\end{array}\right)$\\
$1^{+}$ &$\left(\begin{array}{cccccc}-69.2&-1.9&30.2&6.8&25.6&-1.6\\-1.9&14.0&8.0&25.6&0.0&-49.8\\30.2&8.0&7.1&-1.6&-49.8&0.0\\6.8&25.6&-1.6&-2.4&-0.8&12.1\\25.6&0.0&-49.8&-0.8&-28.0&3.2\\-1.6&-49.8&0.0&12.1&3.2&-14.1\end{array}\right)$&$\left(\begin{array}{c}-106.4\\-61.4\\57.7\\40.7\\-28.8\\5.5\end{array}\right)$&$\left[\begin{array}{cccccc}\{0.73,0.10,-0.43,-0.09,-0.51,0.10\}\\\{-0.13,0.58,0.05,-0.38,0.09,0.70\}\\\{0.05,0.75,0.34,0.22,-0.20,-0.49\}\\\{0.08,-0.31,0.76,-0.13,-0.51,0.22\}\\\{0.66,-0.01,0.35,-0.10,0.65,-0.03\}\\\{-0.10,-0.03,-0.04,-0.88,-0.04,-0.46\}\end{array}\right]$&$\left(\begin{array}{c}8936.5\\8981.5\\9100.6\\9083.6\\9014.1\\9048.4\end{array}\right)$&$\left(\begin{array}{c}8297.5\\8342.5\\8461.6\\8444.6\\8375.1\\8409.4\end{array}\right)$&$\left(\begin{array}{c}8365.9\\8410.8\\8529.9\\8512.9\\8443.4\\8477.7\end{array}\right)$\\
$0^{+}$ &$\left(\begin{array}{cccc}-25.9&-5.5&13.6&86.2\\-5.5&-63.2&86.2&0.0\\13.6&86.2&-127.9&-13.9\\86.2&0.0&-13.9&31.6\end{array}\right)$&$\left(\begin{array}{c}-190.9\\94.0\\-85.4\\-3.0\end{array}\right)$&$\left[\begin{array}{cccc}\{-0.14,-0.56,0.81,0.11\}\\\{0.58,-0.04,-0.03,0.81\}\\\{0.79,-0.17,0.10,-0.57\}\\\{0.10,0.81,0.57,-0.01\}\end{array}\right]$&$\left(\begin{array}{c}8852.0\\9136.9\\8957.5\\9039.9\end{array}\right)$&$\left(\begin{array}{c}8213.0\\8497.9\\8318.5\\8400.9\end{array}\right)$&$\left(\begin{array}{c}8281.4\\8566.2\\8386.9\\8469.2\end{array}\right)$\\
\hline
\end{tabular}
\end{table*}

\begin{figure*}[!h]
\begin{tabular}{lcr}
\resizebox{0.45\textwidth}{!}{\includegraphics{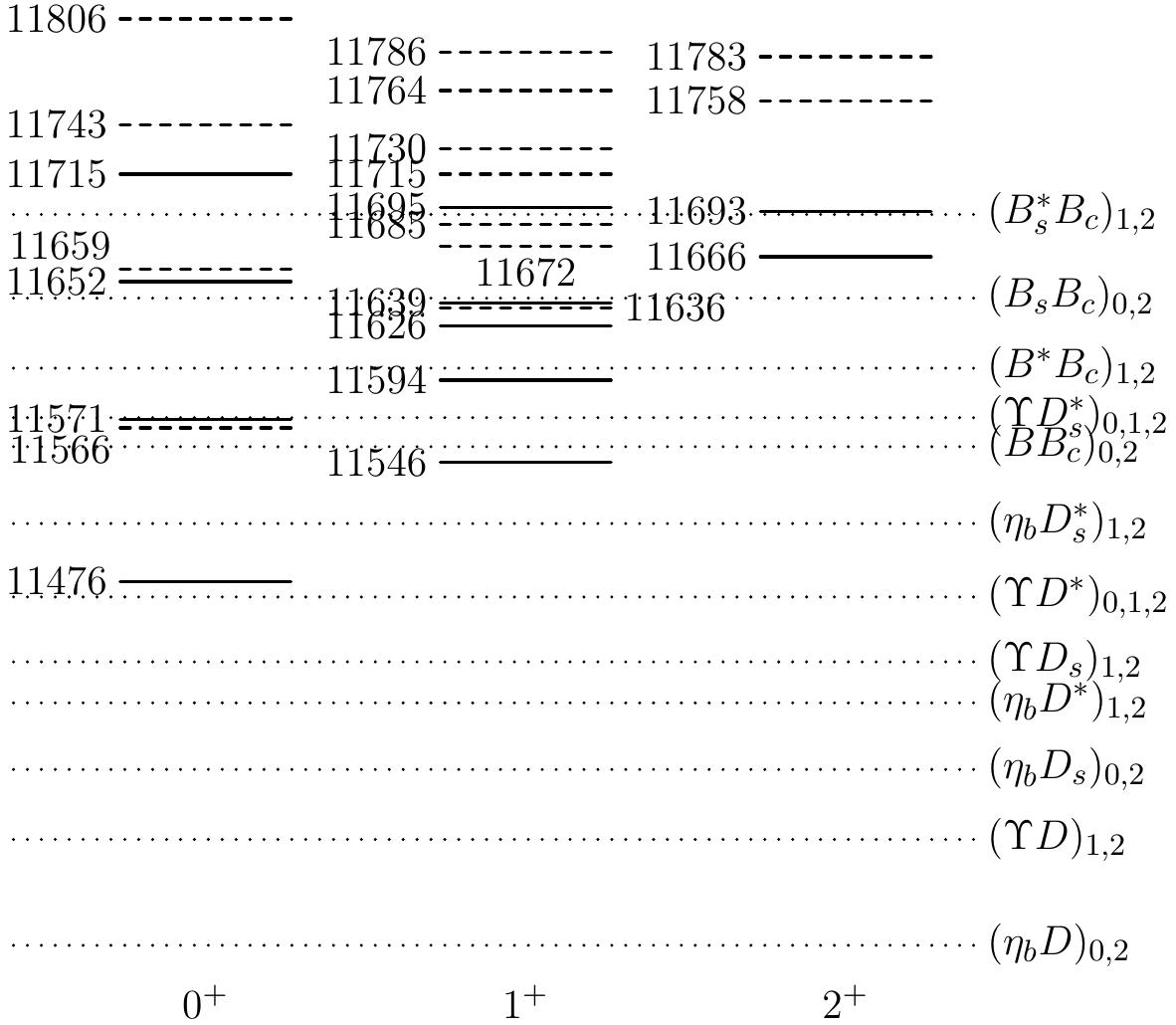}}&\qquad&
\resizebox{0.45\textwidth}{!}{\includegraphics{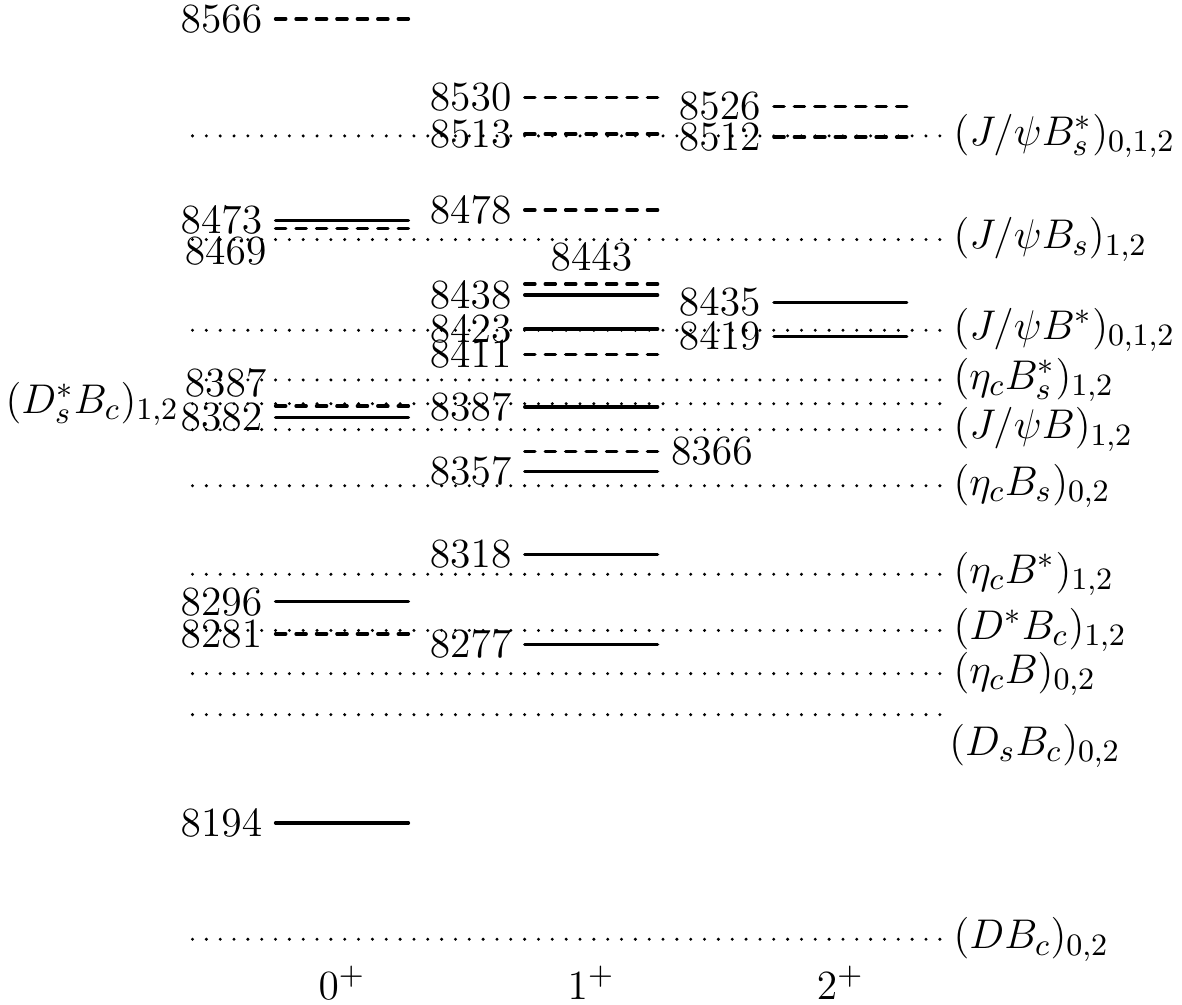}}
\end{tabular}
\caption{Proposed $bc\bar{b}\bar{q}$ (left) and $bc\bar{c}\bar{q}$
(right) tetraquark states. The solid (dashed) line corresponds to
the case $q=u,d$ ($q=s$). The dotted line indicates various
meson-meson thresholds. When a number in the subscript of a
meson-meson state is equal to the spin of an initial state, the
decay for the initial state into that meson-meson channel through
$S$- or $D$-wave is allowed. The masses are given in units of
MeV.}\label{QpQQp-q}
\end{figure*}

The $bc\bar{b}\bar{q}$ and $bc\bar{c}\bar{q}$ are also hidden-bottom
and hidden-charm systems respectively. Their features are different
from the states in the last subsection. The former case corresponds
to the excited $D$ and $D_s$ mesons with much higher masses than the
$cc\bar{c}\bar{q}$. The latter case corresponds to the excited
$\bar{B}$ and $\bar{B}_s$ mesons with lower masses than the
$bb\bar{b}\bar{q}$. Now the first two heavy quarks are different in
the flavor space and there is no constraint from the Pauli
principle. Therefore, the number of allowed states is doubled. There
are two types of meson-meson threshold one may compare to,
$(c\bar{b})(b\bar{q})$ and $(b\bar{b})(c\bar{q})$
($(b\bar{c})(c\bar{q})$ and $(c\bar{c})(b\bar{q})$) for the
$bc\bar{b}\bar{q}$ ($bc\bar{c}\bar{q}$) case. We use both of them in
estimating the tetraquark masses. The number of the possible strong
decay channels is also bigger than in the previous cases. We show
the numerical results in Tabs. \ref{result:bcbq} and
\ref{result:bccq} for the $bc\bar{b}\bar{q}$ and $bc\bar{c}\bar{q}$
systems, respectively. At present, we cannot determine the accurate
values of the masses without solving the bound state problem. One
needs further study to answer which set of masses is more physical.
The rough positions for these states are given in Fig.
\ref{QpQQp-q}, where we use the masses estimated with the $BB_c$,
$B_sB_c$, $BJ/\psi$, and $B_sJ/\psi$ thresholds.

When discussing the decay patterns, we do not include possible final
states containing $B_c^*$. First, we focus on the $bc\bar{b}\bar{q}$
case. For the states with $J^P=0^+$, possible $S$-wave channels are
$\eta_bD$, $\Upsilon D^*$, $\bar{B}B_c$, $\eta_bD_s$, $\Upsilon
D_s^*$, and $\bar{B}_sB_c$. For the states with $J^P=1^+$, possible
channels are $\Upsilon D$, $\eta_bD^*$, $\Upsilon D^*$,
$\bar{B}^*B_c$, $\Upsilon D_s$, $\eta_bD_s^*$, $\Upsilon D_s^*$, and
$\bar{B}_s^*B_c$. For the $J^P=2^+$ states, possible channels are
just $\Upsilon D^*$ and $\Upsilon D_s^*$. Secondly, we take a look
at the $bc\bar{c}\bar{q}$ case. The possible $S$-wave channels for
$J^P=0^+$ are $\eta_c\bar{B}$, $J/\psi\bar{B}^*$, $B_cD$,
$\eta_c\bar{B}_s$, $J/\psi\bar{B}_s^*$, and $B_cD_s$. For $J^P=1^+$,
the channels are $J/\psi\bar{B}$, $\eta_c\bar{B}^*$,
$J/\psi\bar{B}^*$, $B_cD^*$, $J/\psi\bar{B}_s$, $\eta_c\bar{B}_s^*$,
$J/\psi\bar{B}_s^*$, and $B_cD_s^*$. For $J^P=2^+$, the channels are
$J/\psi\bar{B}^*$ and $J/\psi\bar{B}_s^*$. More channels are
possible if the $D$-wave decay is considered. Whether the channels
are open or not is easy to judge from Fig. \ref{QpQQp-q}.

\subsection{Numerical results in triquark-$q$ configuration}

In section \ref{sec2}, we have found that the diquark-antidiquark
configuration and the triquark-antiquark configuration give
identical results for the cases $J=2$ and $J=0$ because the
flavor-color-spin wave functions are the same. For the $J=1$ case,
their explicit spin wave functions are different, which results from
the coupling order in the spin space. The resulting color-spin bases
are different and the CMI matrices are not equal if the color-spin
mixing is not considered. After the diagonalization for the matrix
$\langle H_{CM}\rangle$ is performed, one finds that the results in
the two configurations are also equal. We show in Tab.
\ref{result:tri-q} the diagonal of the matrix and its eigenvalues in
the triquark-antiquark configuration for the case $J=1$. It is
obvious that the tetraquark spectra in these two configurations are
the same from the comparison with previous results. By comparing the
numbers in the two columns of Tab. \ref{result:tri-q}, one
understands the importance of the mixing effect in the
triquark-antiquark configuration.

\begin{table*}[!h]
\caption{The diagonal of the matrix $\langle H_{CM}\rangle$ and its
eigenvalues for the various systems in the case $J^P=1^+$ in the
triquark-antiquark configuration. The values are given in units of
MeV.}\label{result:tri-q} \centering
\begin{tabular}{ccc}\hline
System&$diag\Big(\langle H_{CM}\rangle\Big)$&Eigenvalues\\
$cc\bar{c}\bar{q}$& $(-19.3,-5.8,15.9)$&$(-72.8,69.4,-5.7)$\\
$cc\bar{c}\bar{s}$&$(-22.0,-6.3,15.1)$&$(-76.0,67.1,-4.3)$\\
$bb\bar{b}\bar{q}$&$(0.4,-5.2,9.9)$&$(32.9,-27.0,-0.8)$\\
$bb\bar{b}\bar{s}$&$(-0.1,-4.7,10.0)$&$(34.2,-26.9,-2.1)$\\
$cc\bar{b}\bar{q}$&$(-12.6, 7.3, 19.5)$&$(70.7, -60.6, 4.0)$\\
$cc\bar{b}\bar{s}$&$(-12.2, 7.4, 19.6)$&$(71.1, -60.3, 4.0)$\\
$bb\bar{c}\bar{q}$&$(-10.6, -9.7, 6.3)$&$(-44.8, 26.7, 4.1)$\\
$bb\bar{c}\bar{s}$&$(-14.1, -9.9, 5.5)$&$(-48.9, 25.5, 4.8)$ \\
$bc\bar{b}\bar{q}$&$(-29.7, -25.6, 11.5, -8.3, -1.1, -22.9)$&$(-95.6, 53.7, -47.5, 31.0, -15.6, -2.2)$\\
$bc\bar{b}\bar{s}$&$(-31.1, -25.2, 11.6, -8.3, -0.8, -23.2)$&$(-95.7, 54.0, -47.0, 31.5, -17.3, -2.5)$\\
$bc\bar{c}\bar{q}$&$(-15.7, -40.4, 7.9, -17.1, -9.9, -15.7)$&$(-105.0, -63.5, 56.7, 41.0, -25.0, 4.9)$\\
$bc\bar{c}\bar{s}$&$(-15.5,-39.7,7.1,-20.2,-10.2,-14.1)$&$(-106.4,-61.4, 57.7,40.7,-28.8, 5.5)$\\
\hline
\end{tabular}
\end{table*}

\section{Discussions}\label{sec4}

In this work, we have considered both the diquark-antidiquark
[$(QQ)(\bar{Q}\bar{q})$] and triquark-antiquark
[$(QQ\bar{Q})\bar{q}$] configurations and obtained the same
numerical results ($Q=c,b$, $q=u,d,s$). We notice that one
does not need to distinguish the configurations for a compact
$QQ\bar{Q}\bar{q}$ system once the mixing between different
color-spin states is considered.

The role of the color-spin mixing is different for the $J=2$, $J=1$,
and $J=0$ cases in a specific system. For all the discussed systems,
the $J^P=0^+$ states get the largest mass gap from the mixing effect
and all the highest and lowest states are scalar. In our
calculation, the largest mass gaps for the $cc\bar{c}\bar{n}$
($cc\bar{c}\bar{s}$), $bb\bar{b}\bar{n}$ ($bb\bar{b}\bar{s}$),
$cc\bar{b}\bar{n}$ ($cc\bar{b}\bar{s}$), $bb\bar{c}\bar{n}$
($bb\bar{c}\bar{s}$), $bc\bar{b}\bar{n}$ ($bc\bar{b}\bar{s}$) and
$bc\bar{c}\bar{n}$ ($bc\bar{c}\bar{s}$) systems are 247 (248) MeV,
103 (107) MeV, 206 (206) MeV, 112 (116) MeV, 239 (240) MeV and 279
(285) MeV, respectively.

We have used two parameter schemes to estimate the masses of the
compact $QQ\bar{Q}\bar{q}$ tetraquarks. In the effective quark mass
scheme, our results are only theoretical upper limits. In the
reference threshold method, the rough masses probably are close to
the physical ones. We collect the rough masses for the systems in
Table \ref{mesons}, where the tetraquark states without constraint
from the Pauli principle and those with exotic flavor are labeled.
The results are preliminary since the calculation does not involve
dynamics. More studies are needed to clarify the mass spectrum of
the $QQ\bar{Q}\bar{q}$ systems. Whether there exist possible stable
$QQ\bar{Q}\bar{q}$ tetraquarks also needs dynamical investigations.
It is easy to find the possible rearrangement decay patterns from
Figs. \ref{QQQ-q}, \ref{QQQp-q}, and \ref{QpQQp-q}.

We have used an oversimplified model to calculate the mass
splittings which are determined by the effective coupling constants
and the color-spin structures. To extract the values of the coupling
constants, we have assumed that they are equal to those in the
conventional hadrons. Since the effective couplings are actually
related to the orbital wave functions which are not necessarily the
same for different hadrons, one has to explore whether this extended
application is appropriate or not in the future. In addition, since
the contributions from the kinetic term, color Coulomb term, and
confinement term are equivalently incorporated in the effective
quark masses, one has to employ an improved model to get more
information about the tetraquark states. Further investigations are
definitely needed to understand the uncertainty in the present
model.

The studied tetraquark systems involve dominantly gluon-exchange
interactions. Besides the phenomenological models relying on the
one-gluon exchange potential and a confinement function, other
non-perturbative methods of QCD, such as lattice simulation, QCD sum
rule, and effective field theories, are also appropriate tools to
investigate the properties of the triply-heavy tetraquark states.
For the explicitly exotic $bb\bar{c}\bar{q}$ and $cc\bar{b}\bar{q}$
states, the lattice QCD calculation does not suffer from the
annihilation effect and the simulation is relatively easier. Such
studies with various approaches will definitely help us understand
whether the genuine tetraquark states exist or not. 

\section{Conclusions}

There is a long history of searching for multiquark states. In
recent years, many exotic XYZ mesons are observed and good
tetraquark candidates have been found. However, it is difficult to
distinguish the compact multiquark picture from the molecular
picture once two or more light quarks are involved in the states. In
other heavy quark cases, no appropriate binding mechanisms exist for
loosely bound molecules and the identification of compact
multiquarks is possible. The triply-heavy tetraquark systems provide
us an opportunity to identify genuine multiquark states. Their
binding force is dominantly provided by the gluon exchange
interactions. Besides the explicitly exotic $cc\bar{b}\bar{q}$ and
$bb\bar{c}\bar{q}$ states, the hidden exotic mesons (excited $D$
($D_s$) and $\bar{B}$ ($\bar{B}_s$) with much higher masses) are
also easy to be identified as genuine tetraquark states. 

We have calculated the mass splittings and estimated the rough
masses of the $QQ\bar{Q}\bar{q}$ tetraquarks where $q=u,d,s$ and
$Q=c$ or $b$. We list their decay patterns as shown in the figures
\ref{QQQ-q}, \ref{QQQp-q}, and \ref{QpQQp-q} which may be helpful to
experimental search. If such a high mass $c\bar{q}$-like state could
be observed, its tetraquark nature is easy to be identified.
Moreover, such a state should be accompanied by many partner states.
We want to emphasize that the derived mass splittings should be
reliable, although the present simple chromomagnetic interaction
model cannot give accurate predictions of the tetraquark masses.
However, one can get the masses of its partner states using the mass
splittings derived in the present work once a $QQ\bar{Q}\bar{q}$
tetraquark is observed in the future. Hopefully these intriguing
states can be searched for at LHC. The $cc\bar c\bar q$ may also be
produced at BELLE2.

\begin{table}[!h]\centering
\caption{Comparison for the masses of different meson systems with
only one light antiquark. The symbol (*) means that the system is
not constrained by the Pauli principle and the number of mesons is
doubled comparing to the states constrained by the Pauli principle.
The symbol (\$) indicates explicitly exotic tetraquark
states.}\label{mesons}
\begin{tabular}{cc|cc}\hline
System& Mass (GeV) & System & Mass (GeV)\\
&&$bb\bar{b}\bar{q}$&$\sim14.7$\\
(*)$cb\bar{b}\bar{q}$&$\sim 11.5$ & (\$)$bb\bar{c}\bar{q}$&$\sim 11.6$ \\
(\$)$cc\bar{b}\bar{q}$&$\sim 8.2$ & (*)$bc\bar{c}\bar{q}$&$\sim 8.2$ \\
$cc\bar{c}\bar{q}$&$\sim5.0$ & $b\bar{q}$&$\sim 5.3$ \\
$c\bar{q}$&$\sim 2.0$ \\\hline
\end{tabular}
\end{table}

\section*{Acknowledgments}

This project is supported by National Natural Science Foundation of
China under Grants No. 11175073. No. 11275115, No. 11222547, No.
11575008, No. 11261130311, the Fundamental Research Funds for the
Central Universities and 973 program. XL is also supported by the
National Program for Support of Young Top-notch Professionals.

\end{document}